\documentclass[letterpaper,english,reprint,aps,prl,superscriptaddress,showpacs, longbibliography]{revtex4-1}
\usepackage[latin9]{inputenc}
\usepackage{color}
\usepackage{amsmath}
\usepackage{amssymb}
\usepackage{graphicx}
\usepackage{algorithm}
\usepackage{algpseudocode}
\usepackage{algorithmicx}

\makeatletter

\pdfpageheight\paperheight
\pdfpagewidth\paperwidth

\usepackage{babel}
\usepackage[colorlinks,
            linkcolor=red,
            anchorcolor=black,
            citecolor=blue
            ]{hyperref}

\makeatother

\usepackage{babel}
\begin{document}
\title{Deep learning enhanced Rydberg multifrequency microwave recognition}
\author{Zong-Kai Liu}
\affiliation{Key Laboratory of Quantum Information, University of Science and Technology
of China, Hefei, Anhui 230026, China}
\affiliation{Synergetic Innovation Center of Quantum Information and Quantum Physics,
University of Science and Technology of China, Hefei, Anhui 230026,
China}
\author{Li-Hua Zhang}
\affiliation{Key Laboratory of Quantum Information, University of Science and Technology
of China, Hefei, Anhui 230026, China}
\affiliation{Synergetic Innovation Center of Quantum Information and Quantum Physics,
University of Science and Technology of China, Hefei, Anhui 230026,
China}
\author{Bang Liu}
\affiliation{Key Laboratory of Quantum Information, University of Science and Technology
of China, Hefei, Anhui 230026, China}
\affiliation{Synergetic Innovation Center of Quantum Information and Quantum Physics,
University of Science and Technology of China, Hefei, Anhui 230026,
China}
\author{Zheng-Yuan Zhang}
\affiliation{Key Laboratory of Quantum Information, University of Science and Technology
of China, Hefei, Anhui 230026, China}
\affiliation{Synergetic Innovation Center of Quantum Information and Quantum Physics,
University of Science and Technology of China, Hefei, Anhui 230026,
China}
\author{Guang-Can Guo}
\affiliation{Key Laboratory of Quantum Information, University of Science and Technology
of China, Hefei, Anhui 230026, China}
\affiliation{Synergetic Innovation Center of Quantum Information and Quantum Physics,
University of Science and Technology of China, Hefei, Anhui 230026,
China}
\author{Dong-Sheng Ding}
\email{dds@ustc.edu.cn}

\affiliation{Key Laboratory of Quantum Information, University of Science and Technology
of China, Hefei, Anhui 230026, China}
\affiliation{Synergetic Innovation Center of Quantum Information and Quantum Physics,
University of Science and Technology of China, Hefei, Anhui 230026,
China}
\author{Bao-Sen Shi}
\email{drshi@ustc.edu.cn}

\affiliation{Key Laboratory of Quantum Information, University of Science and Technology
of China, Hefei, Anhui 230026, China}
\affiliation{Synergetic Innovation Center of Quantum Information and Quantum Physics,
University of Science and Technology of China, Hefei, Anhui 230026,
China}
\date{\today}

\maketitle

\textbf{Recognition of multifrequency microwave (MW) electric fields is challenging because of the complex interference of multifrequency fields in practical applications. Rydberg atom-based measurements for multifrequency MW electric fields is promising in MW radar and MW communications. However, Rydberg atoms are sensitive not only to the MW signal but also to noise from atomic collisions and the environment, meaning that solution of the governing Lindblad master equation of light-atom interactions is complicated by the inclusion of noise and high-order terms. Here, we solve these problems by combining Rydberg atoms with deep learning model, demonstrating that this model uses the sensitivity of the Rydberg atoms while also reducing the impact of noise without solving the master equation. As a proof-of-principle demonstration, the deep learning enhanced Rydberg receiver allows direct decoding of the frequency-division multiplexed (FDM) signal. This type of sensing technology is expected to benefit Rydberg-based MW fields sensing and communication.}

\textbf{\textsf{\textbf{\textsf{Introduction}}}}

The strong interaction between Rydberg atoms and microwave (MW) fields that results from their high polarizability
means that the Rydberg atom is a candidate medium for MW fields measurement, e.g.
using electromagnetically induced absorption \citep{EIA2020}, electromagnetically
induced transparency (EIT) \citep{RevModPhys.77.633,AT2017} and the
Autler\textendash Townes effect \citep{sedlacek2012microwave,AT1955,AT2017,EITvsAT2010}.
The amplitudes \citep{meyer2018digital,jiao2019atom,mixer2019,jing2020atomic}, phases
\citep{simons2019rydberg, jing2020atomic} and frequencies \citep{mixer2019, jing2020atomic} of
MW fields could then be measured with high sensitivity. Based on this measurement sensitivity
for MW fields, the Rydberg atom has been used in communications
\citep{holloway2019detecting,meyer2018digital,jiao2019atom,Holloway2019guitar}
and radar \citep{angle2021} as an atom-based radio receiver.
In the communications field, the Rydberg atom replaces the traditional
antenna with superior performance aspects that include sub-wavelength size,
high sensitivity, system international (SI) traceability to Planck\textquoteright s
constant, high dynamic range, self-calibration and an operating
range that spans from MHz to THz frequencies \citep{bason2010enhanced,Multiple2020,mixer2019,meyer2018digital, jing2020atomic}.
One application is analogue communications, e.g. real-time recording
and reconstruction of audio signals \citep{Holloway2019guitar}. Another application
is digital communications, e.g. phase-shift keying and quadrature
amplitude modulation \citep{holloway2019detecting,meyer2018digital,jiao2019atom}.
The channel capacity of MW-based communications is limited by the standard
quantum limited phase uncertainty \citep{meyer2018digital}. Furthermore,
a continuously-tunable radio-frequency carrier has been realized based
on Rydberg atoms \citep{continuously2019}, thus paving the way for concurrent
multichannel communications. Detection and decoding of multifrequency
MW fields are highly important in communications for acceleration of
information transmission and improved bandwidth efficiency.
Additionally, MW fields recognition enables simultaneous detection of multiple
targets with different velocities from the multifrequency spectrum
induced by the Doppler effect. However, because of the sensitivity of
Rydberg atoms, the noise is superimposed on the message, meaning that the
message cannot be recovered efficiently. Additionally, it is difficult to generalize
and scale the band-pass filters to enable demultiplexing of multifrequency
signals with more carriers \citep{Multiple2020}.

To solve these problems, we use a deep learning model for its
accurate signal prediction capability and its outstanding ability to recognize complex
information from noisy data without use of complex circuits. The deep learning model updates the weights via backpropagation and then extracts
features from massive data without human intervention or prior knowledge of physics and the experimental system. Because of these
advantages, physicists have constructed complex neural networks to
complete numerous tasks, including far-field subwavelength acoustic imaging
\citep{PhysRevX.10.031029}, value estimation of a stochastic
magnetic field \citep{khanahmadi2020timedependent}, vortex light recognition \citep{PhysRevLett.124.160401,PhysRevLett.123.183902}, demultiplexing
of an orbital angular momentum beam \citep{ISI:000399872100026,MLOAM} and automatic control of
experiments \citep{Wigley_2016,Deep_mot2018,BayesianGHZ2020,Mills_2020Cool,Cartpoles2020,Bukov_2018Rl}.

\begin{figure*}[t]
\includegraphics[width=2\columnwidth]{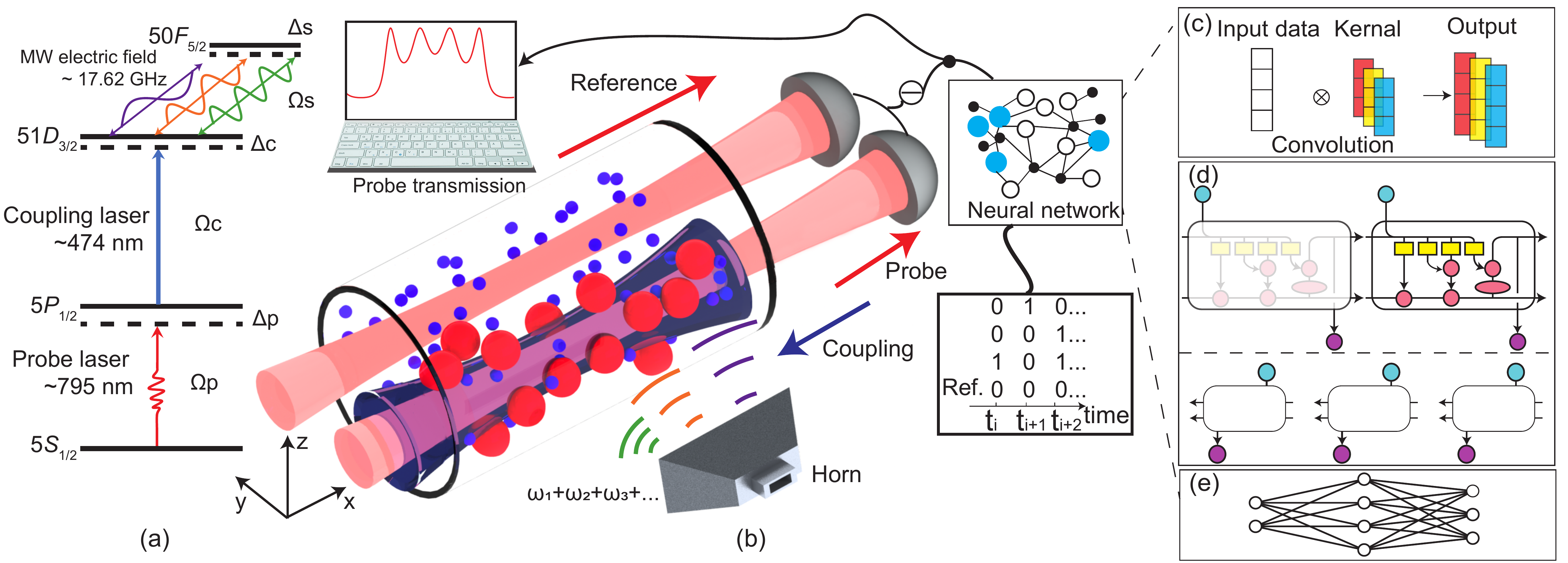}\caption{\textbf{Illustration of the setup.} (a) Overview of experimental energy diagram. Probe and coupling
laser beams excite the atoms at ground state $\left|5S_{1/2}\right\rangle $
to the Rydberg state $\left|51D_{3/2}\right\rangle $. Multifrequency
microwave (MW) electric fields couple the Rydberg states $\left|51D_{3/2}\right\rangle $
and $\left|50F_{5/2}\right\rangle $. (b) Schematic of Rydberg atom-based
antenna and mixer interacting with multifrequency signals. A 795 nm
laser beam is split into two beams, which then propagate in parallel through a heated Rb
cell (length: 10 cm, temperature: $44.6^{\circ}\mathrm{C}$, atomic density: $9.0\times10^{10}$ $\mathrm{cm^{-3}}$ \citep{vsibalic2017arc}). One is the probe beam, which counterpropagates
with the coupling laser beam exciting atoms to Rydberg states to reduce Doppler broadening. The other is the
reference beam, which does not counterpropagate with the coupling laser beam.
The beams are detected using a differencing photodetector (DD) to obtain the probe
transmission spectrum (inset). Multifrequency MW fields transmitted
by a horn are applied to the atoms, with a radiated direction that is perpendicular
to the laser beam propagation direction. The multifrequency
MW fields are modulated using a phase signal such that the phase differences
between the reference bin and the other bins carry the messages. The probe
transmission spectrum is fed into a well-trained neural network to
retrieve the variations of the phases with time. (c-e) Schematics of the neural
network. The network consists of (c) a one-dimensional convolution layer, (d) a bi-directional
long-short term memory layer and (e) a dense layer; for further details about these layers, see the Methods section.}

\label{setup}
\end{figure*}

Here, we demonstrate a deep learning enhanced Rydberg receiver for
frequency-division multiplexed digital communication. In our experiment,
the Rydberg atoms act as a sensitive antenna and a mixer to receive
multifrequency MW signals and extract information \citep{mixer2019,simons2019rydberg,holloway2019detecting}.
The modulated signal frequency is reduced from several gigahertz
to several kilohertz via the interaction between the Rydberg atoms
and the microwaves, thus allowing the information to be extracted using simple
apparatus. These interference signals are then fed into a well-trained
deep learning model to retrieve the messages. The deep learning model
extracts the multifrequency microwave signal phases, even
without knowing anything about the Lindblad master equation,
which describes the interactions between atoms and light beams
in an open system theoretically. The solution of the master equation is often complex
because the higher-order terms and the noises from the environment
and from among the atoms are taken into consideration. However, the deep learning
model is robust to the noise because of its generalization ability, which
takes advantage of the sensitivity of the Rydberg atoms while also reducing
the impact of the noise that results from this sensitivity. Our deep learning
model is scalable, allowing it to recognize the information carried by more than
20 MW bins. Additionally, when the training is complete, the deep learning
model extracts the phases more rapidly than via direct solution of the master
equation.

\begin{figure*}
\includegraphics[width=2\columnwidth]{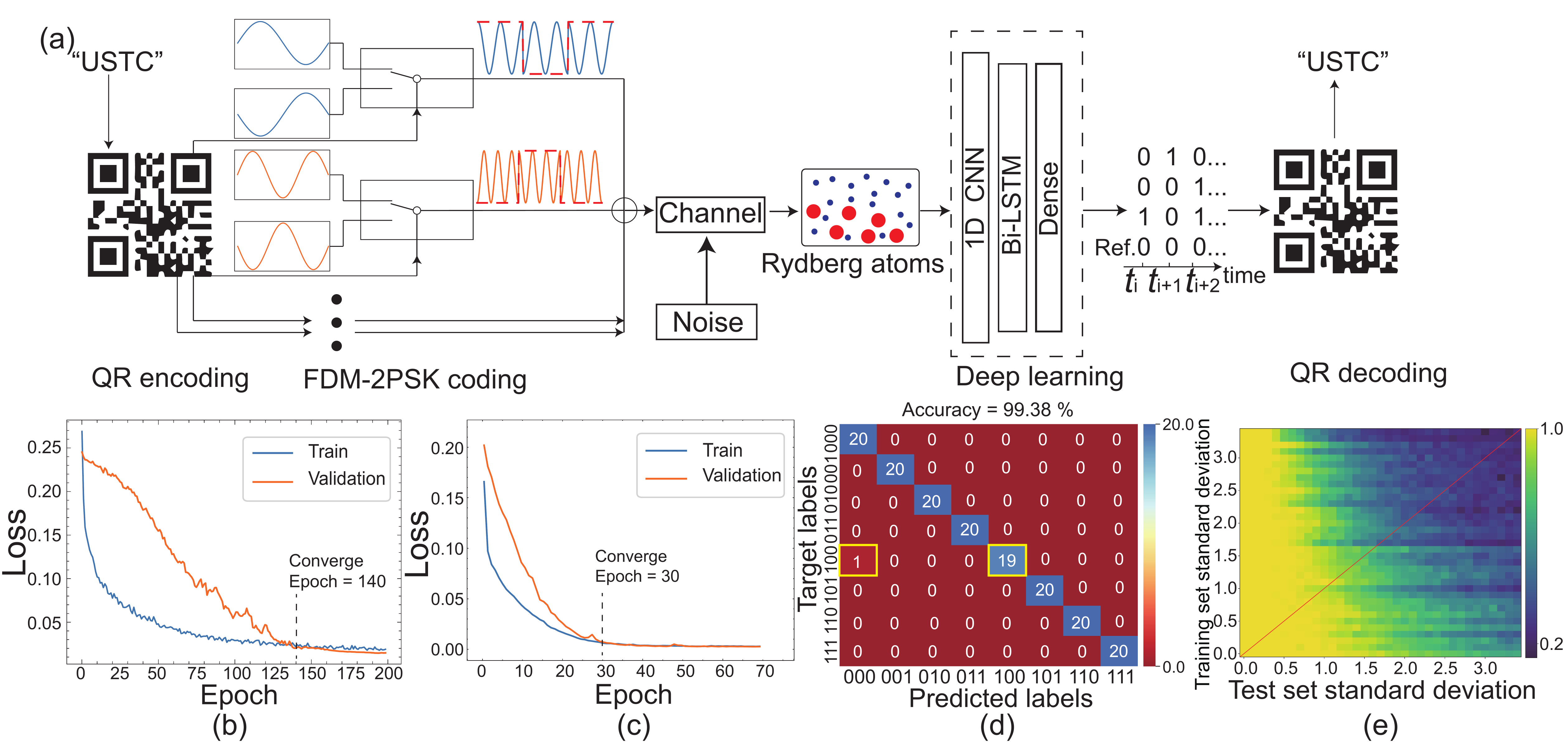}

\caption{\textbf{Flow chat of recognizing the multifrequency MWs and results.}  (a) Process of encoding and decoding frequency-division multiplexed binary phase-shift keying (FDM-2PSK) signal with Rydberg
atoms and the deep learning model.  (b-c) Loss curve evolution
with epochs for the training set (blue) and the validation set (orange)
during training with different training data and validation data.
The training data sizes are (b) 393 and (c) 1194. The validation data sizes are (b) 131 and (c) 398; more details about the data set split are presented in the Methods section. The loss curves for training and validation
converge at (b) 140, and (c) 30 epochs, where an epoch is a time
unit during which the model iterates once over the complete data set; see Methods section.
(d) Confusion matrix for a test set (the number of the testing
set is 160 and the labels are uniformly distributed) after training
in case (c). The accuracy reaches 99.38\% after a 70-epoch training period. (e) Deep learning model accuracy on the noisy test set after training on the noisy training set. The x- and y-axes represent the standard deviations of the additional white noise added to the test set and the training set, respectively. The colorbar represents the accuracy of the model on the noisy test set. The results were obtained by averaging five sets of predictions. The diagonal (red line) indicates the accuracy of the model on a test set in which the noise distribution is the same as that of the training set; more details about the noise are shown in supplementary materials.}

\label{encoding_and_decoding}
\end{figure*}

\textbf{\textsf{Results}}

\textbf{Setup.} We adapt a two-photon Rydberg-EIT scheme to excite
atoms from a ground state to a Rydberg state. A probe field drives
the atomic transition $\left|5S_{1/2},F=2\right\rangle \rightarrow\left|5P_{1/2},F'=3\right\rangle $
and a coupling light couples the transition $\left|5P_{1/2},F'=3\right\rangle \rightarrow\left|51D_{3/2}\right\rangle $
in rubidium 85, as shown in Fig.~\ref{setup}(a). Multifrequency
MW fields drive a radio-frequency (RF) transition between the two different Rydberg
states $\left|51D_{3/2}\right\rangle $ and $\left|50F_{5/2}\right\rangle $.
The energy difference between these states is $17.62\,\mathrm{GHz}$.
The multifrequency MW fields consist of multiple MW bins (more
than three bins) with frequency differences of several kilohertz from
the resonance frequency. The amplitudes, frequencies and phases of the multiple MW bins
can be adjusted individually (further details are provided in the Methods
section). The detunings of the probe, coupling and MW fields are $\Delta_{\text{p}}$,
$\Delta_{\text{c}}$ and $\Delta_{\text{s}}$, respectively. The Rabi frequencies of the probe,
coupling and MW fields are $\Omega_{\text{p}}$, $\Omega_{\text{c}}$, and $\Omega_{\text{s}}$,
respectively. The experimental setup is depicted in Fig.~\ref{setup}(b).
We use MW fields to drive the Rydberg states constantly, producing modulated EIT spectra, i.e. the probe transmission
spectra, as shown in the inset of Fig.~\ref{setup}(b). The phases of the
 MW fields correlate with the modulated EIT spectra and can be recovered
from these spectra with the aid of deep learning. Specifically, the
probe transmission spectra are fed into a well-trained deep learning
model that consists of a one-dimensional convolution layer (1D CNN),
a bi-directional long-short term memory layer (Bi-LSTM) and a dense layer
to extract the phases of the MW fields. Figure~ \ref{setup}(c-e) show
these components of the neural network (further details are presented in the
Methods section). Finally, the bin phases are recovered
and the data are read out.

\begin{figure*}
\includegraphics[width=1.8\columnwidth]{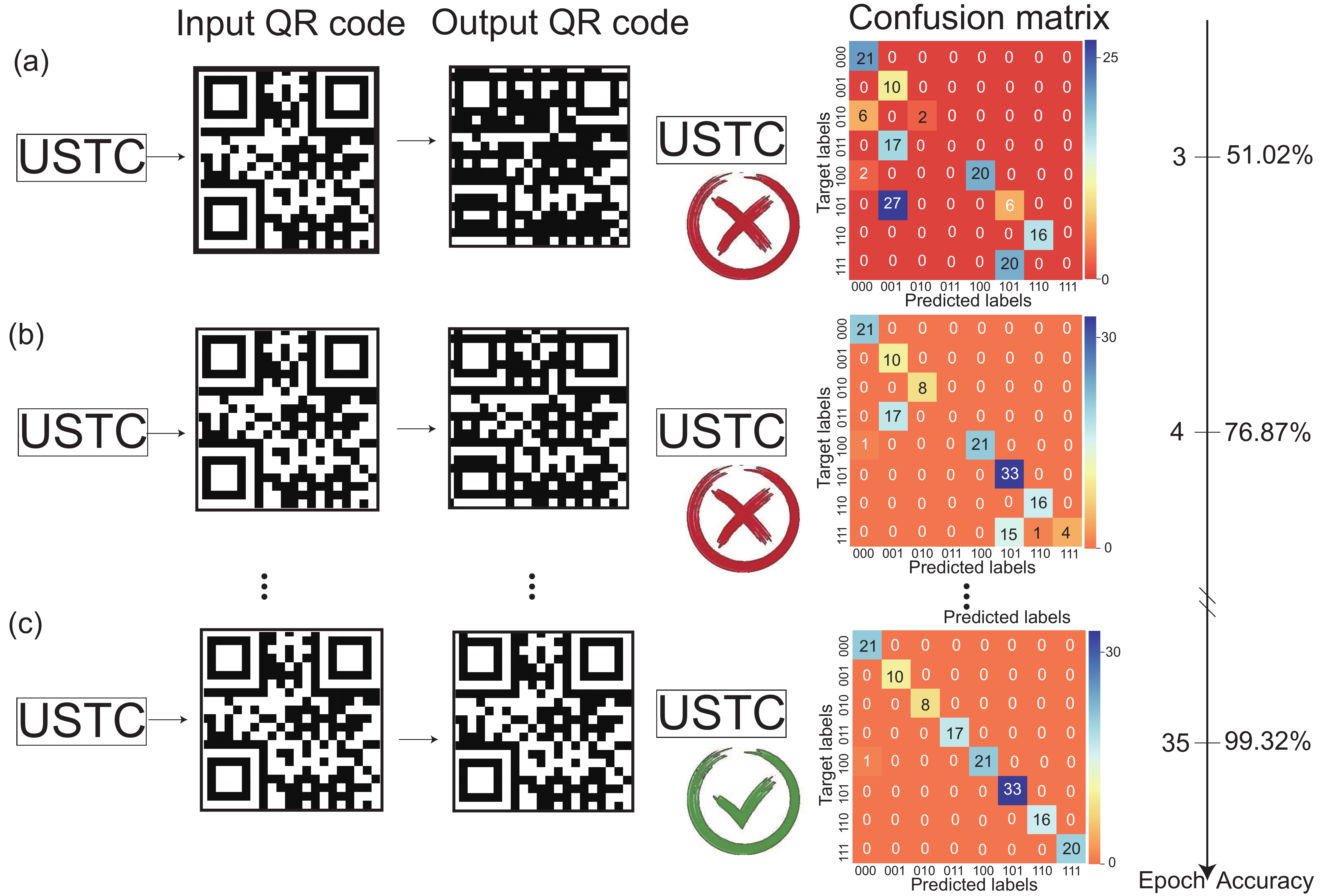}\caption{\textbf{Reconstruction of the QR code after different epochs of training
of the deep learning model.} After (a) 3 epochs, (b) 4 epochs and
(c) 35 epochs of training, the accuracies of our deep learning model
are $51.02\%,\,76.87\%,\,99.32\%$, respectively, as determined from the
confusion matrices (last column). The $y$-axis is the ground truth
and the $x$-axis represents the predictions of the deep learning model.
The colour bar and the numbers indicate the counts. If the prediction is
correct, then the corresponding diagonal term in the confusion matrix increases
by 1. If incorrect, then the nondiagonal element increases by 1.}

\label{confusion}
\end{figure*}

\textbf{FDM signal encoding and receiving.} In the experiments, we use
a four-bin frequency-division multiplexing (FDM) MW signal for demonstration,
where one of the four MW bins is used as the reference bin. The relative phase
differences between the reference bin and the other bins are modulated
by the message signal. Specifically, for the four-bin MW signal, $E=A_{1}\cos[(\omega_{0}+\omega_{1})t+\varphi_{1}]+A_{2}\cos[(\omega_{0}+\omega_{2})t+\varphi_{2}]+A_{3}\cos[(\omega_{0}+\omega_{3})t+\varphi_{3}]+A_{4}\cos[(\omega_{0}+\omega_{4})t+\varphi_{4}],$ where
$\omega_{0}$ is the resonant frequency, $\omega_{1,2,3}$ are the
relative frequencies, the carrier frequencies are $2\pi(\omega_{0}+\omega_{1})=17.62\,\mathrm{GHz}-3\,\mathrm{kHz}$,
$2\pi(\omega_{0}+\omega_{2})=17.62\,\mathrm{GHz}-1\,\mathrm{kHz}$,
$2\pi(\omega_{0}+\omega_{3})=17.62\,\mathrm{GHz}+1\,\mathrm{kHz}$
and $2\pi(\omega_{0}+\omega_{4})=17.62\,\mathrm{GHz}+3\,\mathrm{kHz}$,
the frequency difference between two frequency-adjacent bins is $\Delta f=2\,\mathrm{kHz}$
and the message signal is $\varphi_{1,2,3}=0\,\mathrm{or}\,\pi$, standing
for 3 bits (0 or 1), and the reference phase is $\varphi_{4}=0$ (which remains unchanged).
The phase list $\left(\varphi_{1},\,\varphi_{2},\,\varphi_{3},\,\varphi_{4}\right)$
is a bit string for time $t_{0}$. By varying the phase of $\varphi_{1,2,3}$
with time, we then obtain the FDM signal for binary phase-shift keying
(2PSK). Additionally, the amplitudes of the four bins are $0.1A_{4}=A_{1,2,3}$
to solve the problem that results from the nonlinearity of the atom,
where the probe transmission spectra of two different bit strings,
e.g. $(0,\,0,\,\pi,\,0)$ and $(0,\,\pi,\,0,\,0)$, are the same (further
details are presented in the Methods section). By increasing
the frequency difference $\Delta f$, we can obtain higher information
transmission rates. For four bins with $\Delta f=2\,\mathrm{kHz}$, the
information transmission rate is $n_{b}\times\Delta f=(4-1)\times2\times10^{3}\,\mathrm{bps}=6\,\mathrm{kbps}$,
where $n_{b}$ is the number of bits. In the experiments, disturbances
originate from the environment and atomic collisions. Because of the sensitivity
of Rydberg atoms to MW fields, the resulting noise submerges our signal. To
use the sensitivity of the Rydberg atoms and simultaneously
minimize the effects of noise, the deep learning model is used
to extract the relative phases $\left(\varphi_{1},\,\varphi_{2},\,\varphi_{3}\right)$.

\textbf{Deep learning.} To improve the robustness and
speed of our receiver, we use a deep learning model to decode
the probe transmission signal. The complete encoding and decoding process
is illustrated in Fig.~\ref{encoding_and_decoding}(a). The Rydberg
antenna receives the FDM-2PSK signal and down-converts this signal
into the probe transmission spectrum. The information is then retrieved
from the spectrum using the deep learning model. The precondition is
that the different bit strings correspond to distinct probe spectra; this
is resolved by setting $0.1A_{4}=A_{1,2,3}$, as discussed earlier. Then,
we combine the 1D CNN layer, the Bi-LSTM layer and the dense layer
to form the deep learning model (See the Methods
section for further details) \citep{chollet2015keras,scikit_learn}. One of the reasons
for using the 1D CNN layer and the Bi-LSTM layer is that the data sequences are
long, which means that prediction of the phases $\overrightarrow{\varphi}=\left(\varphi_{1},\,\varphi_{2},\,\varphi_{3},\,0\right)$
from the spectrum is a regression task and requires a long-term
memory for our model. Another reason is to combine the
convolution layer's speed with the sequential sensitivity of the Bi-LSTM layer
\citep{deep_learning_with_python}. The input sequence is first processed
by the 1D CNN to extract the features, meaning that a long sequence is
converted into a shorter sequence with higher-order features. This process is visualized to show how the deep learning model treats the transmission spectrum; more details are presented in the supplementary materials. The
shorter sequence is then fed into the Bi-LSTM layer and resized by the dense
layer to match the label size (see the Methods section for further details). Specifically, the probe spectrum $\mathbf{T}=\left\{ T_{0},\,T_{\tau},\,T_{2\tau},\,\cdots,\,T_{i\cdot\tau}\cdots,\,T_{N\cdot\tau}\right\} $
and the corresponding phases $\overrightarrow{\varphi}=\left(\varphi_{1},\,\varphi_{2},\,\varphi_{3},\,\varphi_{4}=0\right)$
are collected to form the data set, where $T_{i\cdot\tau}$ is the
i-th data point of a probe spectrum and the fourth bit $\varphi_{4}=0$
is the reference bit. Both the spectra and the phases are 1D vectors
with dimensions of $N+1$ and 4, respectively. These independent, identically
distributed data \{\{$\mathbf{T}$\}, \{$\overrightarrow{\varphi}$\}\}
are fed into our model as a data set. By shuffling this data set and
splitting it into three sets, i.e. a training set, a validation set and
a test set, we train our model on the training set (feeding both
the waveforms and labels \{\{$\mathbf{T}$\}, \{$\overrightarrow{\varphi}$\}\}),
 validate and test our model on the validation and test sets, respectively (by feeding
waveforms without labels and comparing the predictions with ground
truth labels). The validation set is used to determine whether there is
either overfitting or underfitting during training. Finally, the performance
(i.e. accuracy) of the model is estimated by predicting the test set.

\begin{figure}[htbp]
\includegraphics[width=0.9\columnwidth]{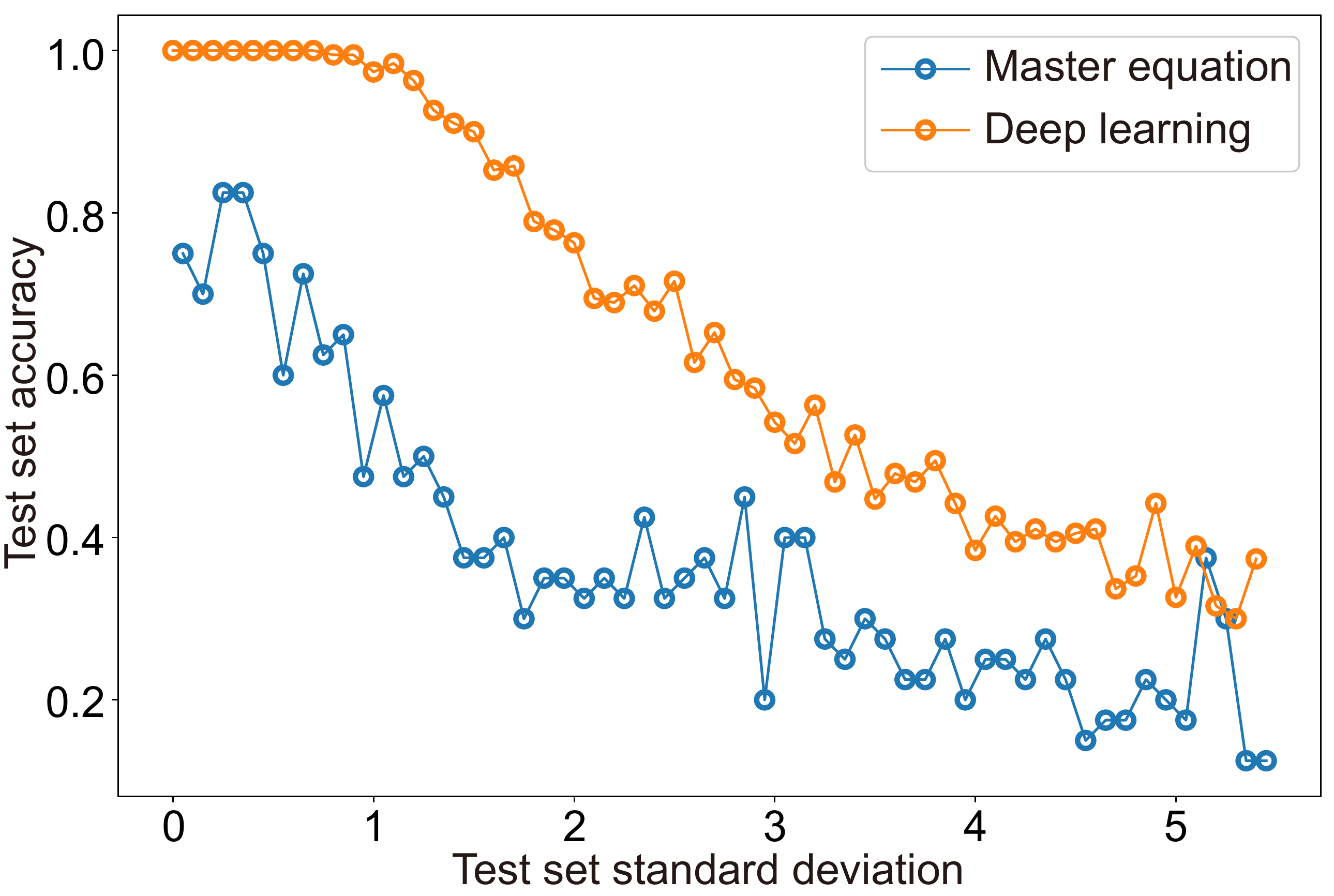}\caption{\textbf{Prediction accuracy of deep learning model and the master equation on the noisy test set.} The noise is white noise with mean $\mu=0$ and a standard deviation $\sigma$. The noise is added quantitatively using different $\sigma$ values. Before and after addition of the noise, the data are scaled between 0 and 1 using their maximum and minimum. The deep learning model is trained in a training set without additional noise. In addition, we do not involve noise spectrum when solving the master equation. Each point is obtained after averaging of five predictions on the noisy data. }
\label{p}
\end{figure}

The performance of our deep learning model is affected by the training epochs and the training
and validation set sizes. The training curves on different training sets
and validation sets are shown in  Figs.~\ref{encoding_and_decoding}(b-c).
Initially, our model performs well on the training set only, implying
overfitting. The curves then converge (dashed line) and our model
performs well on both the training set and the validation set. The
sudden jump in the loss curve in Fig.~\ref{encoding_and_decoding}(c)
is caused by the change in the learning rate (further details are presented
in the Methods section). Use of more training and validation
data causes the curves to converge more quickly. The deep learning model performs well after these few-sample training.
In Fig.~\ref{encoding_and_decoding}(d),
we show a confusion matrix for prediction of a uniformly
distributed test set, which demonstrates accuracy of 99.38\%.

The ``noise'' shown in Fig.~\ref{encoding_and_decoding}(a) refers to two kinds of noises. One comes from atoms and the external environment (systematic noise). The other comes from the noise added on purpose (additional noise). The systematic noise cannot be adjusted quantitatively and is discussed with its noise spectrum  in the supplementary materials. Because the noise on the data set is independent and is distributed identically (i.i.d.), i.e., the entire data set is shuffled before being split into the training and test sets, the systematic noise pattern is almost the same in both the training set and the test set. The deep learning model has already learned the systematic noise pattern during the training process, which is one of the major advantages of use of deep learning against systematic noise.  However, there is a case where the noise is not i.i.d. (i.e., the case where a specific noise occurs during testing only). This problem can be solved by online learning and addition of prior knowledge as new features into the data, e.g., data for the temperature, the weather, and other factors \citep{sahoo2017online}.  Here for simplicity, we talk about the i.i.d. case only and add the white noise with a mean $\mu$ and a standard deviation $\sigma$. We ignore the 1/f noise in this case because it decays rapidly in the low frequency range and the signal with which it would interfere is located within the $2\sim200$ kHz range. The additional noise is added both on the training set and the test set of the deep learning model in Fig.~\ref{encoding_and_decoding}(e), which demonstrates the performance of the deep learning model when used on a data set with biased or unbalanced noise. The results below the red line show the performance of the model after training on a weaker-noise training set when predicting based on a stronger-noise test set, i.e., generalization for a stronger noise case. These results indicate that the deep learning model has the generalization ability required to adapt to stronger noise. In the area above the red line, there is more noise in the training set than in the test set. Theoretically, a small amount of additional noise in the training set will increase the robustness of the deep learning model. However, when the noise increases, it affects the accuracy, which decays rapidly. Next, the well-trained model is used to reconstruct the QR code. In Figs.~\ref{confusion}(a-c), the results and the corresponding
confusion matrices with their epochs are shown. First, the information is
encoded into a QR code. After the code is transmitted, received and
decoded using the Rydberg atoms and the deep learning model, the
information is then reconstructed successfully using the 35-epoch training
model in Fig.~\ref{confusion}(c), but is not reconstructed in parts (a) and (b). The accuracy
is defined by the number of correctly predicted bit strings divided
by the total number of bit strings (147 bit strings). After 35 epochs,
the accuracy reaches $99.32\%$ and the message is reconstructed
successfully from the QR code received.

\textbf{Comparison between deep learning method and the master equation.} In our case, the master equation that we employed is the commonly used one without considering the noise spectrum. The accuracies of the deep learning model and the master equation fitting on noisy data are different. Fig.~\ref{p} shows the accuracies obtained by the two methods. The deep learning model is trained on a training set without additional noise, and tested on a test set with additional white noise whose standard deviation is $\sigma$ (the transmission spectra with noise are given in supplementary materials). Here for simplicity, the data set is composed of the transmission of four MW bins only (one of them is reference bin) and the frequency difference between the adjoin bins is $\Delta f = 2 \mathrm{kHz}$. On the other hand, the result of the master equation is given based on the same test set as that of the deep learning model. The deep learning method outperforms the fitting of the master equation on the noisy data set.

\begin{figure*}[t]
\includegraphics[width=2\columnwidth]{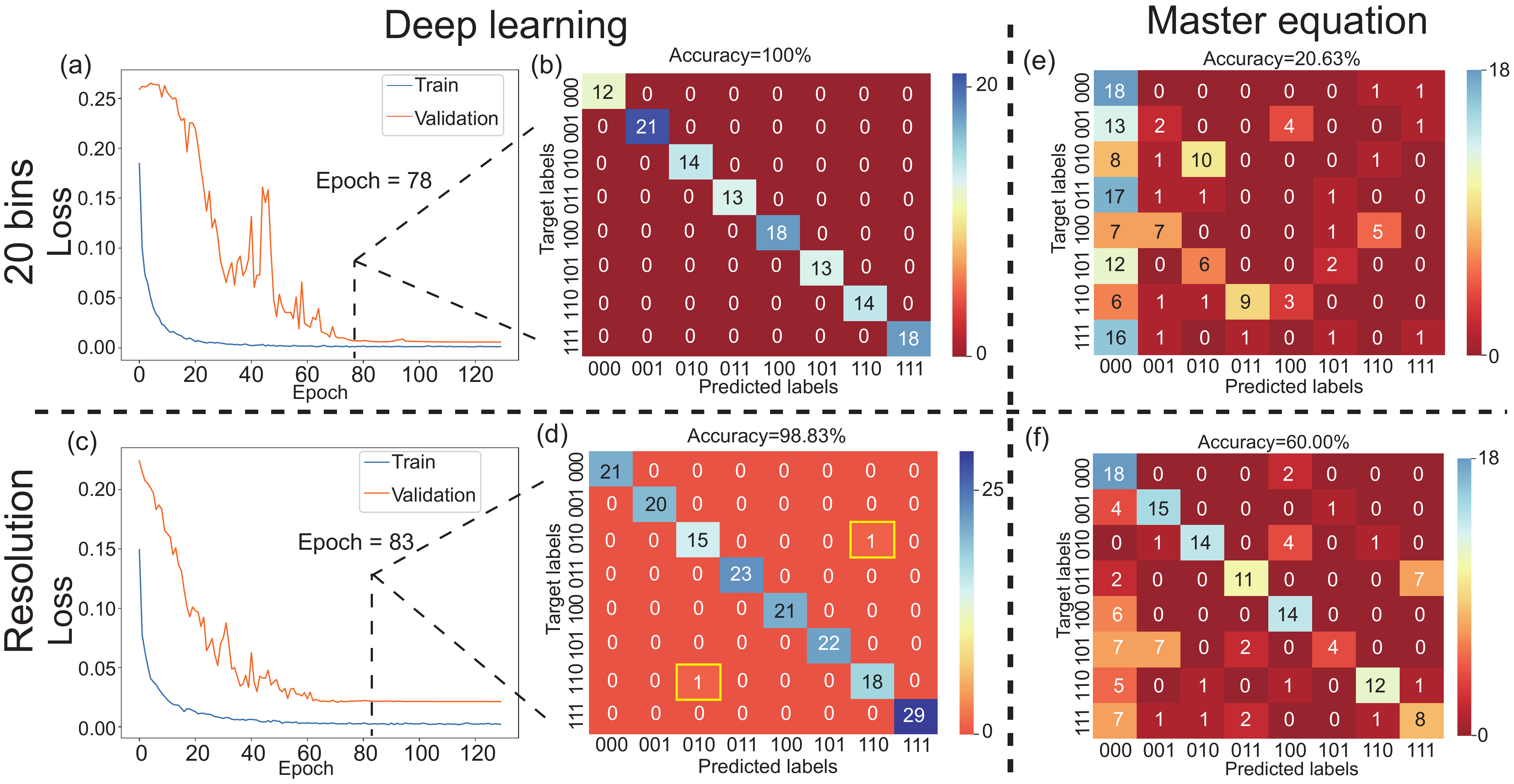} \caption{\textbf{Loss curves and confusion maps of the deep learning method and the master equation fitting method for  MWs with 20 bins and frequency difference $\Delta f$.} (a) Loss vs. epoch curves for training and validation of a 20-bin
carrier with frequency difference $\Delta f=2\,\mathrm{kHz}$. (b)
Confusion matrix for epoch=78 on the test data set. Only the first
3 bits of the 20 bits are selected to carry the messages and all other
bits are 0. The accuracy is 100\% for 123 testing spectra. (c)
Loss vs. epoch curves for training and validation of the four MW bins
(one bin is the reference) with frequency difference $\Delta f=200\,\mathrm{kHz}$.
(d) Confusion matrix for epoch=83 on the test data set. The accuracy
is 98.83\% for 171 testing spectra. (e) Predicted solution
of the master equation without consideration of higher-order terms and noise on
a 20-bin carrier with frequency difference $\Delta f=2\,\mathrm{kHz}$.
The number of spectra is 160, where the spectra were sampled uniformly from eight categories.
The accuracy is 20.63\%. (f) Predicted solution of the master equation
without consideration of higher-order terms and noise on four-bin carrier with
frequency difference $\Delta f=200\,\mathrm{kHz}$. The number of
spectra is 160, which were sampled uniformly from
eight categories. The accuracy is 60.00\%. In (b) and (e), only the first
3 bits of a total of 20 bits (including a reference) are labelled.}
\label{20bins_res}
\end{figure*}

Apart from the robutness to the noise, when the transmission rate is increased by increasing the number
of MW bins or the frequency difference $\Delta f$, the deep learning model performs
well, while it is difficult to retrieve the messages with
high accuracy using the master equation. Specifically, to increase the bandwidth efficiency
and the transmission rate, the number of MW bins used to carry the messages
must be increased, but the information is still recognizable because of the scalability of the
deep learning model. For 20 MW bins, the number of bits is $(20-1)$
with one reference bit, giving a $\left(20-1\right)\times2\,\mathrm{kbps}=38\,\mathrm{kbps}$
transmission rate. The number of combinations of these bits is $2^{19}$,
which increases exponentially as the number of MW bins increases. Here, for
demonstration purposes, only the first 3 bits of the total of 19 bits carry the
messages and the other bits, including the reference, are set to be 0. To
show how well our model performs, we train, validate and test
the model on this new data set without varying the other parameters, with the exception
of the training epochs of our model. The loss curves for training
and validation are shown in Fig.~\ref{20bins_res}(a). A confusion
matrix for epoch 78 is shown in Fig.~\ref{20bins_res}(b). The
model performs well on this new test set, which was sampled uniformly
from eight categories with accuracy of 100\%. Another method that can be used to increase
the information transmission rate involves increasing the frequency difference.
In our case, the frequency difference is increased from $\Delta f=2\,\mathrm{kHz}$
to $\Delta f=200\,\mathrm{kHz}$. The transmission rate increases
correspondingly, from $(4-1)\times2\,\mathrm{kbps}=6\,\mathrm{kbps}$
to $(4-1)\times200\,\mathrm{kbps}=0.6\,\mathrm{Mbps}$. To detect
the high-speed signal, the DD bandwidth is increased, which
inevitably leads to increased noise. After the model is trained on this
new data set, the training and validation loss curves are as shown
in Fig.~\ref{20bins_res}(c). A confusion matrix for epoch 83 is
shown in Fig.~\ref{20bins_res}(d). Increasing the number of training epochs
allows the model to perform well on this new data set, with accuracy of 98.83\%
on a uniformly sampled test set.

To compare the performances of the deep learning model and the master equation,
we fitted the probe spectra for 20 bins with a frequency difference $\Delta f=2\,\mathrm{kHz}$
and four bins with a frequency difference $\Delta f=200\,\mathrm{kHz}$
by solving the master equation without considering the higher-order terms
and the effects of noise. In each case, 160 probe spectra were fitted
that were sampled uniformly from every category. The prediction results
are shown in Fig.~\ref{20bins_res}(e) and (f). The prediction
accuracy of the master equation is lower than that of the deep learning model.
In our case, the impact of increasing the number of bins is greater than increasing the DD bandwidth for high-speed signals on the fitting accuracy. The prediction accuracy for a 20-bin carrier
with frequency difference $\Delta f=2\,\mathrm{kHz}$ is 20.63\%, which
is like to the accuracy of guessing, i.e., $1/8$.
This implies that there is a disadvantage that comes from the fitting method itself, i.e., it can easily become trapped by local minima. Some type of prior knowledge is required to overcome this disadvantage, e.g., provision of the initial values of the phases before fitting. In contrast, the deep learning model is data-driven and does not require any prior knowledge.  The local minima problem of deep learning can be overcome using some well-known techniques, including learning rate scheduling and design of a more effective optimizer  \citep{deep_learning_with_python}. Additionally, the accuracy
difference for the 200-kHz-difference MW bins between the deep learning model
and the master equation means that the deep learning model is more robust
to noise.  Furthermore, the prediction time for the master equation is $25\,\mathrm{s}$
per spectrum, while the
time for the deep learning model is $1.6\mathrm{\,ms}$ per spectrum.
The master equation is solved by ``FindFit'' function in
Mathematica 11.1 with both ``AccuracyGoal'' and ``PrecisionGoal'' default, while the deep learning code is written in Python 3.7.6. These codes are run on the same computer with NVIDIA GTX 1650 and  Intel\textregistered $\mathrm{Core}^{\mathrm{TM}}$ i7-9750H.

Another method to decode the signal is available that uses an in-phase and quadrature (I-Q) demodulator or a lock-in amplifier \citep{holloway2019detecting, meyer2018digital}. However, the carrier frequency must be given when decoding the signal in this case. Additionally, for multiple MW bins, numerous bandpass filters are required. The deep learning method is thus much more convenient.

\textbf{\textsf{Discussion}}

We report a work on Rydberg receiver enhanced via deep learning
to detect multifrequency MW fields. The results show that the
deep learning enhanced Rydberg mixer receives and decodes multifrequency
MW fields efficiently; these fields are often difficult to decode using
theoretical methods. Using the deep learning model, the Rydberg receiver is robust
to noise induced by the environment and atomic collisions
and is immune to the distortion that results from the limited bandwidths
of the Rydberg atoms (from dipole-dipole interactions and the EIT pumping
rate, as studied in Ref. \citep{meyer2018digital}) for high-speed
signals ($\Delta f=200\,\mathrm{kHz}$). In addition to increasing the transmission speed
of the signals, further increments in the information transmission rate
are achieved by using more bins, which is feasible because of the scalability
of our model. Besides the transmission rate, this deep learning enhanced Rydberg system promises for use in studies of the channel capacity limitations. Because spectra that are difficult for humans to recognize as a result of noise and distortion are distinguishable when using the deep learning model, Rydberg systems enhanced by deep learning could take steps toward the realization of the capacity limit proposed in the literature Ref.~\citep{cox2018quantum}. To obtain high performance (i.e. high signal-to-noise
ratio, information transmission rate, channel capacity and accuracy), the training epochs
and training set must be extended and enlarged.

In summary, we have demonstrated the advantages
of receiving and decoding multifrequency signals using a deep learning
enhanced Rydberg receiver. In a multifrequency signal receiver,
rather than using multiple band-pass filters, lock-in amplifier \citep{holloway2019detecting, meyer2018digital} and other complex
circuits, signals can be decoded using the extremely sensitive Rydberg
atoms and the deep learning model at high speed and with high accuracy
without solving the Lindblad master equation. One of the advantages of use of the Rydberg atom is that the accuracy of the Rydberg atom approaches the photon shot noise limit \citep{Kumar17}. In principle, the accuracy of the Rydberg atom is higher than that of the classical antenna. According to recent work based on the atomic superheterodyne method, ultrahigh sensitivity can be obtained \citep{jing2020atomic}. However, in this proof-of-principle demonstration, there is considerable room for the optimization required to reach that limit (e.g., stabilization of the laser, narrowing the laser linewidth, and temperature stabilization). The sensitivity of the
Rydberg atoms is a double-edged sword because it also involves noise. The deep
learning model restricts this side effect while taking full advantage of
the Rydberg atoms' sensitivity to the signal. Using the automatic feature extraction processes of
the neural networks, the spectra are classified in a supervised manner.
If the features (e.g. mean value, variance, frequency spectrum) are
extracted manually, the spectra are then clustered by unsupervised
learning methods such as t-distributed stochastic neighbour embedding (t-SNE) or the density-based spatial clustering of applications with noise (DBSCAN) method \citep{scikit_learn}, without
training on the training set. Our work will be useful in fields
including high-precision signal measurement and atomic sensors. Additionally, this decoding ability can be generalized further to decode other signals that are encoded by different encoding protocols, e.g., frequency division multiplexing amplitude shift keying (FDM-ASK), frequency division multiplexing quadrature amplitude modulation (FDM-QAM), and IEEE 802.11ac WLAN standard signals for a 5 GHz carrier. The frequency of carrier to be decoded covers from several hertz to terahertz since for Rydberg atoms to receive MW with different wavelengths, the only part of the system that needs to be tuned is the frequency of the laser, while in classical receivers, the wavelength of the received MW is limited by the size of the antenna \citep{PhysRevApplied.15.014053, Meyer_2020,wade2018terahertz,holloway2014broadband}.  In addition to communications, our
receiver can be used to detect multiple targets from multifrequency
signals caused by the Doppler effect.

\textbf{\textsf{Methods}}

\textbf{Generation and calibration of MW fields.} The MW fields used
in our experiments were synthesized by the signal generator (1465F-V from Ceyear)
and a frequency horn. Each bin in the multifrequency MW field is
tunable in terms of frequency, amplitude and phase. The RF source
operates in the range from DC to 40 GHz. The frequency horn is located close
to the Rb cell. We used an antenna and a spectrum analyzer (4024F from Ceyear) to receive
the MW fields and then calibrated the amplitudes of the MW fields at the centre
of the Rb cell.

The probe transmission spectrum in the time domain when $\Delta_{\text{p}}=0$,
$\Delta_{\text{c}}=0$ and $\Delta_{\text{s}}=0$ reflects the interference among
the multifrequency MW bins, which results from the beat frequencies
of the bins that occur through the interaction between the atoms and light. The Rydberg atoms receive
the MW bins by acting as an antenna and a mixer \citep{mixer2019,simons2019rydberg,holloway2019detecting}.
After reception by the atoms, the frequency spectrum of the probe
transmission shows that we can obtain the frequency differential
signal from the probe transmission spectrum. This represents an application
of our atoms to reduce the modulated signal frequency (from
terahertz to kilohertz magnitude), which allows the signal to be received
and decoded using simple apparatus. In our experiment, more than 20
frequency bins can be added to the atoms, for which the dynamic range is greater
than $30\,\mathrm{dBm}$. The amplitudes, phases and frequencies
of these bins can be tuned individually. When the bandwidth is increased to
detect an increasing frequency difference $\Delta f$ signal, more
noise is involved, but this noise is suppressed by the deep learning
model. In other words, the signal can be recognized using the deep learning
model when the information transmission rate is increased by raising
the frequency difference $\Delta f$. These bins are used to send
FDM-PSK signals
in the ``FDM signal encoding and receiving '' section of the main text.

\textbf{Master equation.} The Lindblad master equation is given as
follows: $\text{d}\rho/\text{d}t=-i\left[H,\rho\right]/\hbar+L/\hbar$, where
$\rho$ is the density matrix of the atomic ensemble and $H=\sum_{k}H[\rho^{(k)}]$
is the atom-light interaction Hamiltonian when summed over all the single-atom
Hamiltonians using the rotating wave approximation. This Hamiltonian has the following
matrix form:

\begin{equation}
H=\hbar\left(\begin{array}{cccc}
0 & -\frac{\Omega_{\text{p}}}{2} & 0 & 0\\
-\frac{\text{\ensuremath{\Omega_{\text{p}}} }}{2} & \Delta_{\text{p}} & -\frac{\Omega_{\text{c}}}{2} & 0\\
0 & -\frac{\text{\ensuremath{\Omega_{\text{c}}}}}{2} & \text{\ensuremath{\Delta_{\text{c}}}}+\text{\ensuremath{\Delta_{\text{p}}}} & -\frac{\Omega_{\text{s}}(t)}{2}\\
0 & 0 & -\frac{\Omega_{\text{s}}(t)}{2} & \text{\ensuremath{\Delta_{\text{c}}}}+\text{\ensuremath{\Delta_{\text{p}}}}+\Delta_{\text{s}}
\end{array}\right),
\end{equation}
where for the MW signal $E=A_{1}\cos[\left(\omega_{0}+\omega_{1}\right)t+\varphi_{1}]+A_{2}\cos[\left(\omega_{0}+\omega_{2}\right)t+\varphi_{2}]+A_{3}\cos[\left(\omega_{0}+\omega_{3}\right)t+\varphi_{3}]+A_{4}\cos[\left(\omega_{0}+\omega_{4}\right)t+\varphi_{4}]$,
we have the Rabi frequency $\Omega_{\text{s}}(t)=\sqrt{E_{1}^{2}+E_{2}^{2}}$
, where $E_{1}=A_{1}\sin[\omega_{1}t+\varphi_{1}]+A_{2}\sin[\omega_{2}t+\varphi_{2}]+A_{3}\sin[\omega_{3}t+\varphi_{3}]+A_{4}\sin[\omega_{4}t+\varphi_{4}]$
and $E_{2}=A_{1}\cos[\omega_{1}t+\varphi_{1}]+A_{2}\cos[\omega_{2}t+\varphi_{2}]+A_{3}\cos[\omega_{3}t+\varphi_{3}]+A_{4}\cos[\omega_{4}t+\varphi_{4}]$.
The Rabi frequency can be derived as follows:

{\footnotesize{}
\begin{align}
E & =\sum_{i=1}^{4}A_{i}\cos\left[\left(\omega_{0}+\omega_{i}\right)t+\varphi_{i}\right]\nonumber \\
 & =\sqrt{E_{1}^{2}+E_{2}^{2}}\cos\left(\omega_{0}t+\arctan\frac{\sum_{i=1}^{4}A_{i}\sin\left(\omega_{i}t+\varphi_{i}\right)}{\sum_{i=1}^{4}A_{i}\cos\left(\omega_{i}t+\varphi_{i}\right)}\right)
\end{align}
}where the second term (which resonates with the energy levels of the Rydberg atoms)
induces the normal EIT spectrum and the first term modulates
that spectrum. In the interaction between the atoms and the MW fields, the
atoms act as a mixer such that the output signal frequency ($\omega_{1}$,
$\omega_{2}$, $\omega_{3}$) is less than the input signal frequency
($\omega_{0}+\omega_{1}$, $\omega_{0}+\omega_{2}$, $\omega_{0}+\omega_{3}$).
The modulation signal's nonlinearity is reduced by setting the
reference and increasing its amplitude as shown in equation \ref{eq:nonlinear},
which is a precondition for recognition of these phases via deep learning.

\begin{align}
 & \sqrt{E_{1}^{2}+E_{2}^{2}}\nonumber \\
 & \approx A_{4}\sqrt{1+2\sum_{i=1}^{3}\frac{A_{i}}{A_{4}}\cos\left[\left(\omega_{4}-\omega_{i}\right)t+\left(\varphi_{4}-\varphi_{i}\right)\right]}\nonumber \\
 & \approx A_{4}+\sum_{i=1}^{3}A_{i}\cos\left[\left(\omega_{4}-\omega_{i}\right)t+\left(\varphi_{4}-\varphi_{i}\right)\right]\label{eq:nonlinear}
\end{align}
where the condition for the approximations on the second line and the third
line is $A_{4}\gg A_{1,2,3}$.

The Lindblad superoperator $L=\sum_{k}L[\rho^{(k)}]$ is composed
of single-atom superoperators, where $L[\rho^{(k)}]$ represents the Lindbladian
and has the following form: $\frac{L[\rho^{(k)}]}{\hbar}=-\frac{1}{2}\sum_{m}\left(C_{m}^{\dagger}C_{m}\rho+\rho C_{m}^{\dagger}C_{m}\right)+\sum_{m}C_{m}\rho C_{m}^{\dagger}$
where $C_{1}=\sqrt{\Gamma_{\text{e}}}\left|g\right\rangle \left\langle e\right|$,
$C_{2}=\sqrt{\Gamma_{\text{r}}}\left|e\right\rangle \left\langle r\right|$
and $C_{3}=\sqrt{\Gamma_{\text{s}}}\left|r\right\rangle \left\langle s\right|$
are collapse operators that stand for the decays from state $\left|e\right\rangle $
to state $\left|g\right\rangle $, from state $\left|r\right\rangle $
to state $\left|e\right\rangle $ and from state $\left|s\right\rangle $
to state $\left|r\right\rangle $ with rates $\Gamma_{\text{e}}$, $\Gamma_{\text{r}}$
and $\Gamma_{\text{s}}$, respectively. Because we are only concerned with the steady
state here, i.e. $t\rightarrow\infty$, the Lindblad master equation can
be solved using $\text{d}\rho/\text{d}t=0$. The complex susceptibility of the EIT
medium has the form $\chi(v)=(\lvert\mu_{ge}\rvert^{2}/\epsilon_{0}\hbar)\rho_{eg}$
, where $\rho_{eg}$ is the element of density matrix solved using the master
equation. The spectrum of the EIT medium can be obtained from the
susceptibility using $T\sim e^{-{\rm Im}[\chi]}.$

\textbf{Deep learning layers.} Our deep learning model consists
of a 1D CNN layer, a Bi-LSTM layer and a dense layer. The mathematical
sketches for these layers are given as follows.

The 1D CNN layer is illustrated in Fig.~\ref{setup}(c). The input
signal convolutes the kernel in the following form:

\begin{equation}
f\otimes g=\sum_{m=0}^{N-1}f_{m}g_{(n-m)}.
\end{equation}
where $f$ represents the input data, $g$ is the convolution kernel, $m$
is the input data index and $n$ is the kernel index.
The 1D CNN extracts the higher-order features from the input data to reduce
the lengths of the sequences fed into the Bi-LSTM layer. Before flowing
into the Bi-LSTM layer, the data pass through the batch normalization
layer, the ReLU activation layer and the max-pooling layer, in that sequence.
For a mini-batch $\mathcal{B}=\left\{ x_{1\cdots m}\right\} $, the
output from the batch normalization layer is $y_{i}=BN_{\gamma,\beta}(x_{i})$
and the learning parameters are $\gamma$ and $\beta$ \citep{Batch}.
The update rules for the batch normalization layer are:

\begin{eqnarray}
\mu_{\mathcal{B}} & \leftarrow & \frac{1}{m}\sum_{i=1}^{m}x_{i}\label{eq:mean}\\
\sigma_{\mathcal{B}}^{2} & \leftarrow & \frac{1}{m}\sum_{i=1}^{m}\left(x_{i}-\mu_{\mathcal{B}}\right)^{2}\label{eq:varance}\\
\hat{x}_{i} & \leftarrow & \frac{x_{i}-\mu_{\mathcal{B}}}{\sqrt{\sigma_{\mathcal{B}}^{2}+\epsilon}}\label{eq:normalize}\\
y_{i} & \leftarrow & \gamma\hat{x}_{i}+\beta\equiv BN_{\gamma,\beta}(x_{i})\label{eq:scale_shift}
\end{eqnarray}
where equations \ref{eq:mean} and \ref{eq:varance} evaluate the
mean and the variance of the mini-batch, respectively; the data are normalized using
the mean and the variance in equation \ref{eq:normalize} and the
results are then scaled and shifted in equation \ref{eq:scale_shift}.
The training is accelerated using the batch normalization layer and
the overfitting is also weakened by this layer. The output then passes through
the ReLU activation layer. The activation function of this layer is $f_{\mathrm{ReLU}}(x)=\max(x,0)$.
The vanishing gradient problem is diminished by this activation function.
Next, the inputs are downsampled in a max-pooling layer \citep{chollet2015keras}.

\begin{figure}[t]
\includegraphics[width=0.9\columnwidth]{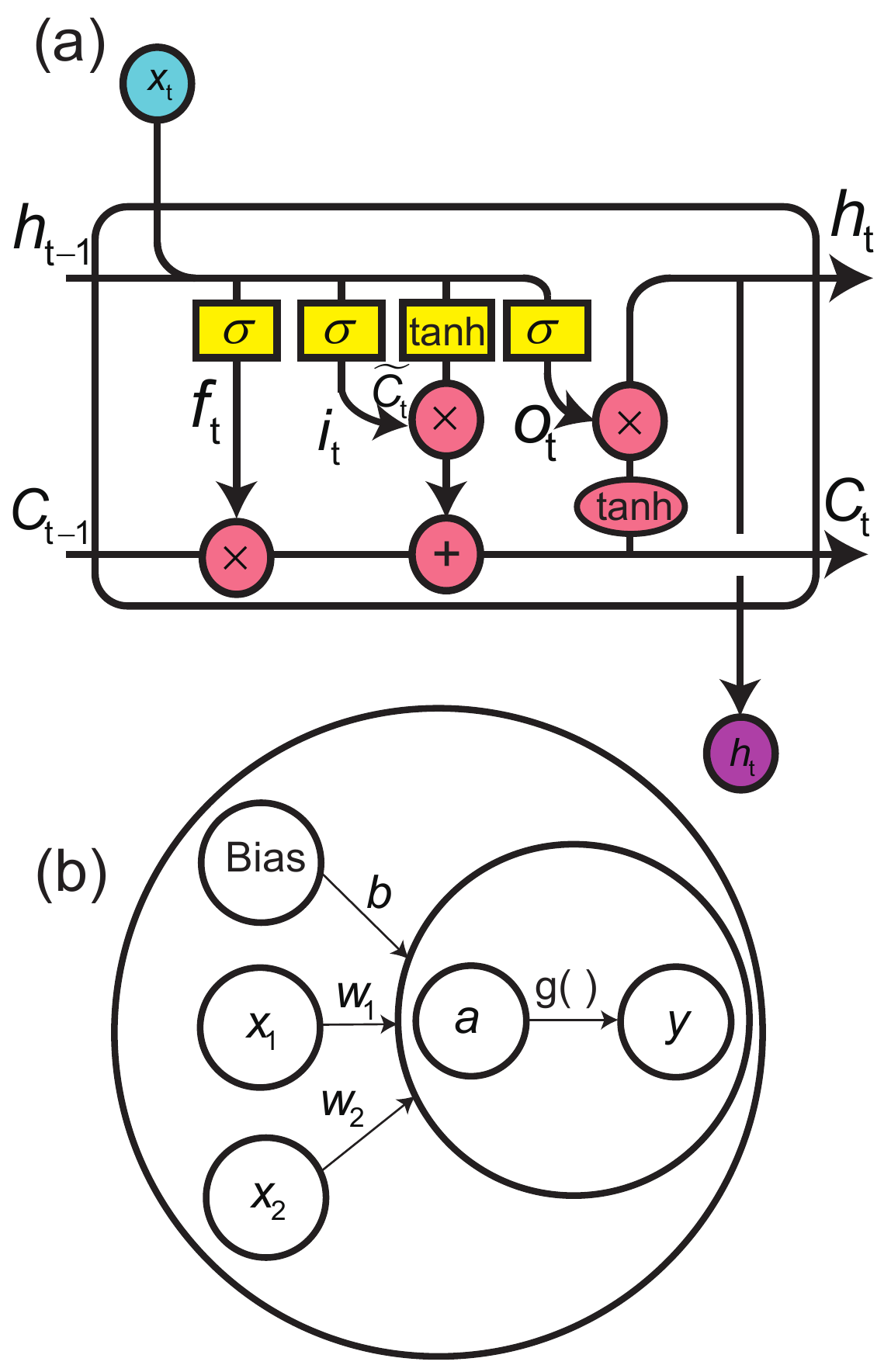}\caption{\textbf{Building blocks for LSTM layer and dense layer.} The cells of a long-short term memory (LSTM) layer and dense layer are presented in (a) and (b), respectively, where
$\sigma$ represents the sigmoid layer and $\tanh$ represents the
tanh layer.}
\label{cell}
\end{figure}

The LSTM layer and an LSTM cell are shown schematically in Fig.~ \ref{setup}(d)
and \ref{cell}(a), respectively. The equations for the LSTM are shown as equations \ref{eq:forget}-\ref{eq:output2}
\citep{deep_learning_with_python,LSTM}. At a time $t$, the input $x_{t}$
and two internal states $C_{t-1}$and $h_{t-1}$ are fed into the
LSTM cell. The first thing to be decided by the LSTM cell is whether or not to forget in
equation \ref{eq:forget}, which outputs a number between 0 and 1
that represents retaining or forgetting. Next, an input gate (equation \ref{eq:input})
decides which values are to be updated from a vector of new candidate values
created using equation \ref{eq:new_value}. The new value is then added
to the cell state and the old value is forgotten in equation \ref{eq:update}.
Finally, the cell decides what to output using equations \ref{eq:output1}
and \ref{eq:output2}.
\begin{eqnarray}
f_{t} & = & \sigma\left(W_{f}\cdot\left[h_{t-1},x_{t}\right]+b_{f}\right)\label{eq:forget}\\
i_{t} & = & \sigma\left(W_{i}\cdot\left[h_{t-1},x_{t}\right]+b_{i}\right)\label{eq:input}\\
\tilde{C}_{t} & = & \tanh\left(W_{C}\cdot\left[h_{t-1},x_{t}\right]+b_{C}\right)\label{eq:new_value}\\
C_{t} & = & f_{t}\times C_{t-1}+i_{t}\times\tilde{C}_{t}\label{eq:update}\\
o_{t} & = & \sigma\left(W_{o}\left[h_{t-1},x_{t}\right]+b_{o}\right)\label{eq:output1}\\
h_{t} & = & o_{t}\times\tanh\left(C_{t}\right)\label{eq:output2}
\end{eqnarray}
where $\sigma(x)=1/(1+e^{-x})$ is the sigmoid function. The sigmoid
and $\tanh$ functions are applied in an element-wise manner. The LSTM is followed
by a time-reversed LSTM to constitute a Bi-LSTM layer that improves the memory
for long sequences.

\begin{figure}[t]
\includegraphics[width=0.9\columnwidth]{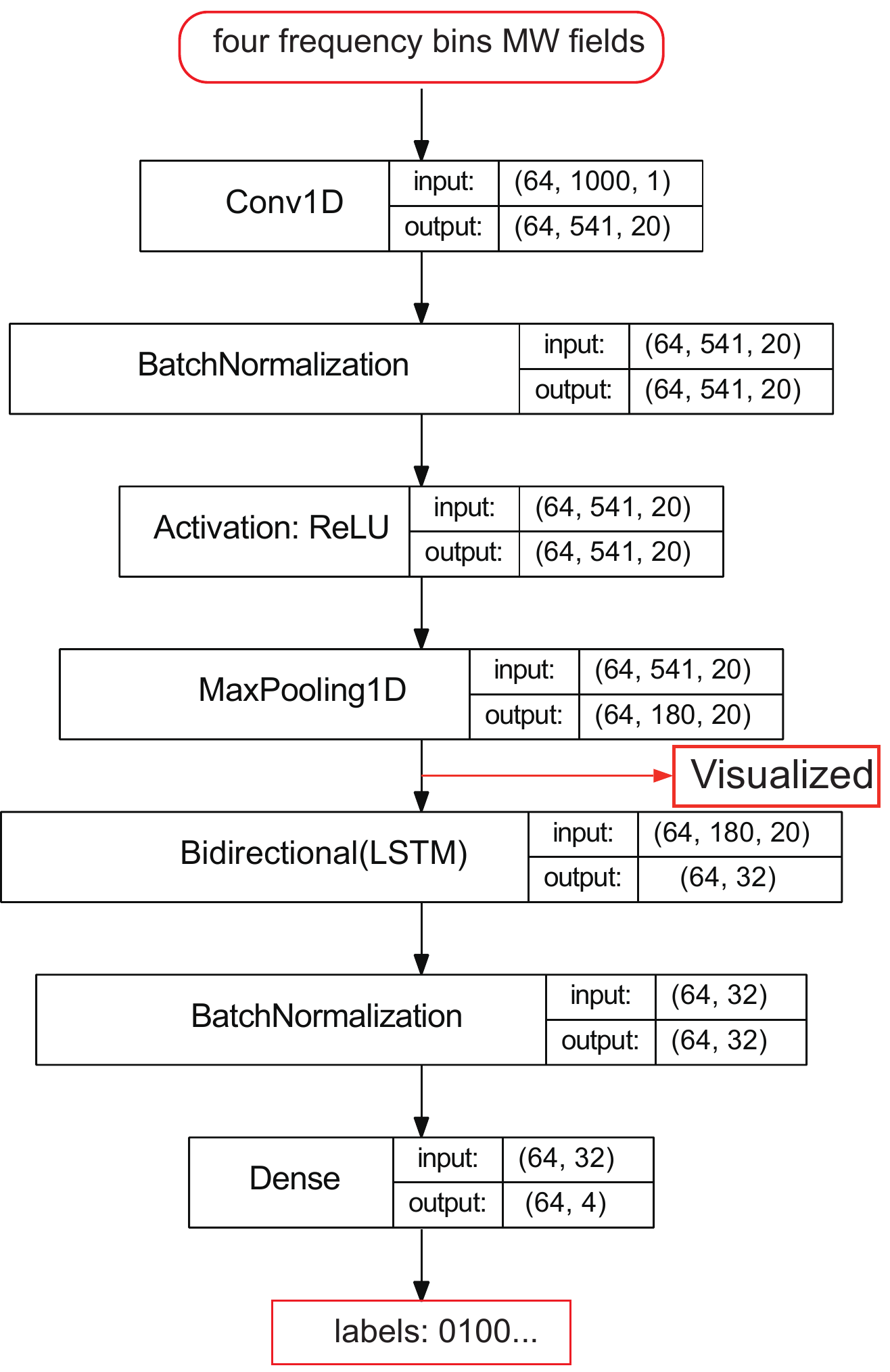}\caption{\textbf{Structure of our deep learning model and size of the data.} The
ReLU activation function is $f(x)=max(0,x)$. After the data output from the max-pooling layer, the visualizations are performed; see supplementary materials for more details.}

\label{Machinelearning}
\end{figure}

The dense layer and a neuron are drawn in Fig.~\ref{setup}(e)
and \ref{cell}(b), respectively, and the corresponding equations are

\begin{eqnarray}
\mathbf{a} & = & \mathbf{w}\cdot\mathbf{x}+b\\
\mathbf{y} & = & g(\mathbf{a})
\end{eqnarray}
where $\mathbf{w}$ is the vector of weights, $b$ is the bias, $x$
represents the input data, $g(\mathbf{a})=1/(1+e^{-\mathbf{a}})$ is the sigmoid
activation function used to limit the output values to between 0 and 1, and
$y$ is the output. The dense layer resizes the shape of the data
obtained from the Bi-LSTM to match the size of the label.

The training consists of both forward and backward propagation. A batch
of probe spectra propagates through the 1D CNN layer, the Bi-LSTM layer
and dense layer during the forward training process. The differentiable
loss function is then calculated. In our case, the differentiable loss
function is the mean squared error (MSE) between the predictions and
the ground truth, which is used widely in the regression task \citep{deep_learning_with_python}.
The equation for the MSE is
\begin{equation}
L_{MSE}=\frac{1}{m\cdot n}\sum_{j=1}^{n}\sum_{i=1}^{m}\left(\varphi_{i,j}-f(T_{i,j})\right)^{2},
\end{equation}
where $m$ is the number of data points in one spectrum, $n$ is the
mini-batch size, $\varphi_{i}$ is the ground truth and $f(T_{i})$
is the model prediction. In backpropagation, the trainable
weights of each layer are updated based on the learning rate and
the derivative of the MSE loss function with respect to the weights to
minimize the loss $L_{MSE}$, such that

\begin{equation}
W\leftarrow W-\eta\frac{\partial L_{MSE}}{\partial W},
\end{equation}
where $\eta$ is the learning rate and $W$ is the trainable weight
for each layer. The weights of each layer are then updated according to the RMSprop optimizer\citep{mini_batch}.

The network is implemented using the Keras 2.3.1 framework on Python
3.6.11 \citep{chollet2015keras}. All weights are initialized with
the Keras default. The hyper-parameters of the deep learning model (including the convolution kernel length, the number of hidden variables
and the learning rate) are tuned using Optuna \citep{optuna_2019}.

\textbf{Deep learning pipeline.} To obtain better fitting results,
the data are scaled based on their maximum and minimum values, i.e. $T_{i}'=(T_{i}-min(T))/(max(T)-min(T))$.
The labels are encoded in dense vectors with four elements rather than
in one-shot encoding vectors to save space \citep{deep_learning_with_python}. Each of these elements is
either 0 or 1, representing the relative phase 0 or $\pi$ of each bin, respectively.

A one-dimensional convolution layer (1D CNN), a bidirectional long
short-term memory layer (Bi-LSTM) and a dense layer are used in
our deep learning model. The deep learning model structure
is shown in Fig.~\ref{Machinelearning}. The data size for the
input layer is given in the form (batch size, length of probe spectrum, number of
features). The batch size is 64 in our case. Because the duration
of the spectrum ranges from $t=0$ to $t=0.999\,\mathrm{ms}$ with a time
difference of $\tau=1\,\mathrm{\mu s}$, the spectrum length is
1000. For a one-dimensional input, the number of features is 1. Therefore,
the data size for the input layer is (64, 1000, 1).

\begin{figure}
    \centering
    \includegraphics[width=0.9\columnwidth]{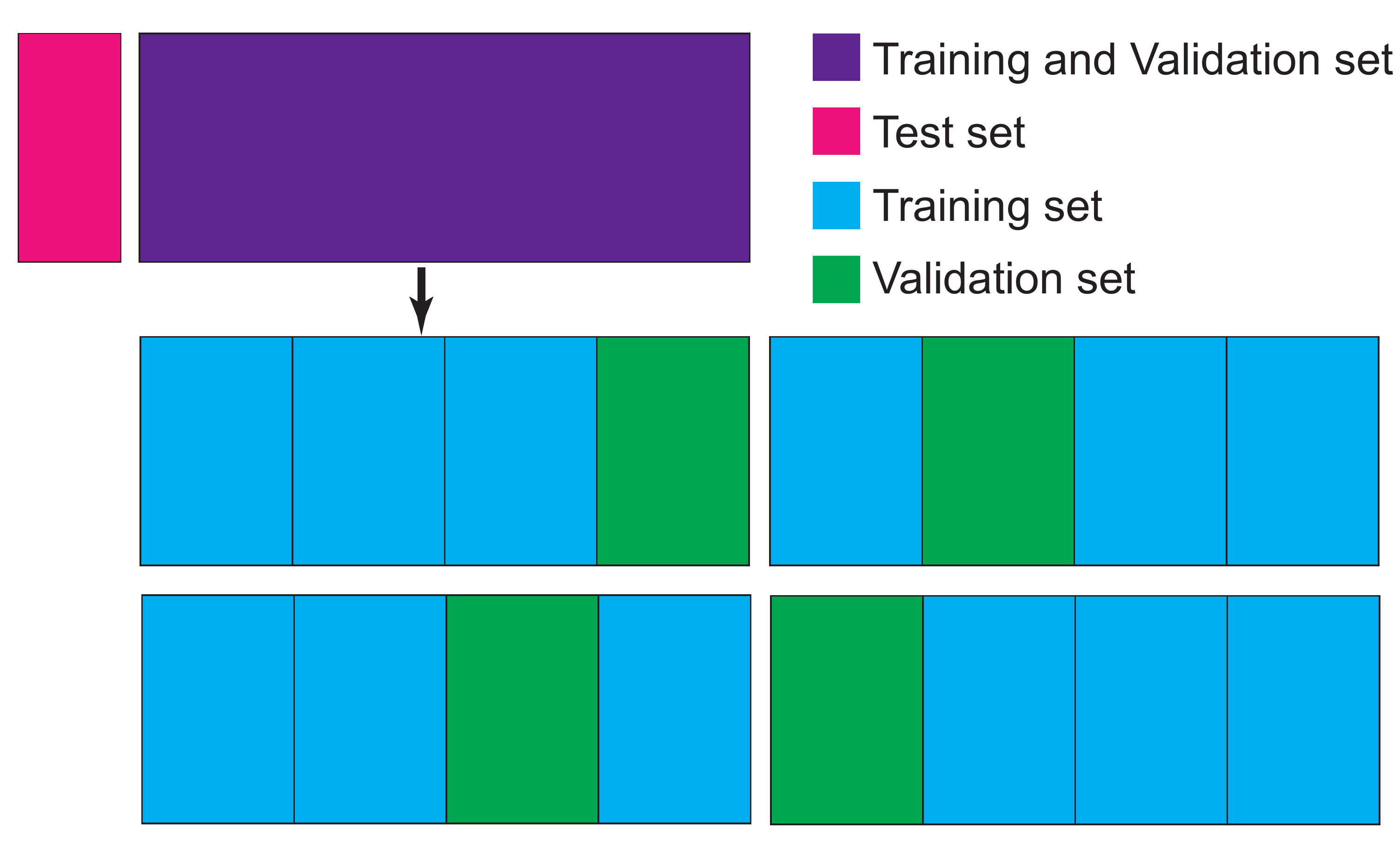}
    \caption{\textbf{Data partition during training, validating, and testing.} First, the data are split into two sets. The first is the test set. The remaining data set is copied four times and is then split into four sets, each with different parts that act as training sets and validation sets for the training of four deep learning models. }
    \label{cross_validation}
\end{figure}

During training of this model, four-fold cross-validation is used to save
the amount of training data.The data set is split as shown in Fig.~\ref{cross_validation}. First, the data set is split into two parts. The first is the test set (red), which remains untouched during training. The second (purple) is used to train the model. In the cross-validation process, the rest data set (purple) is copied four times and is divided equally into four parts each. One of these parts is the validation data set (green) and the others are used as training sets (blue). Four models are trained on the different training sets and validation sets. Then the best model is chosen according to the validation set and is tested on the test set. After splitting, the training set, the validation set, and the test set all remain unchanged. In every epoch, each model iterates the training set only once. There is no new set being taken; instead, the same training set is iterated once each epoch.

 The computational graph is cleared before
each training sequence to prevent leakage of the validation data. Gaussian
noise (where the mean is 0 and the standard deviation is 0.5) is added to the
training data to increase the robustness of the proposed model. In addition, the
learning rate is adjusted during training to jump out of the local minimum,
which results in the jump in Figure \ref{encoding_and_decoding}(c)
in the main text. The initial learning rate is 0.001. If the loss
(mean-square error) of the validation set does not decrease over 10 epochs,
the learning rate is multiplied by 0.1. The RMSprop optimizer is used to
update the weight of each layer during training \citep{mini_batch}.

The bidirectional LSTM layer can be replaced with the well-known self-attention
layer to improve the memory of our proposed model further \citep{vaswani2017attention}.
However, this would require more training time and increased GPU memory. The current model
has been able to meet our requirements to date.

\textbf{Data availability.} The data that support this study are available in Github \citep{Code} (\href{https://github.com/ZongkaiLiu/Deep-learning-enhanced-Rydberg-multifrequency-microwave-recognition}{https://github.com/ZongkaiLiu/Deep-learning-enhanced-Rydberg-multifrequency-microwave-recognition}).

\textbf{Code availability.} The deep learning code and Mathematica code are provided in Github \citep{Code} (\href{https://github.com/ZongkaiLiu/Deep-learning-enhanced-Rydberg-multifrequency-microwave-recognition}{https://github.com/ZongkaiLiu/Deep-learning-enhanced-Rydberg-multifrequency-microwave-recognition}).

\textbf{Acknowledgments.} Zong-Kai Liu gratefully acknowledges the
instructive discussion about deep learning with Yue Chen at the National
Engineering Laboratory for Speech and Language Information Processing and the enlightenment of leraning machine learning during studying with Dr. Lei Gong at Department of Optics and Optical Engineering at USTC.
D.-S.D acknowledges funding from the National Key R\&D Program of China (Grant No. 2017YFA0304800),
the National Natural Science Foundation of China (Grant Nos. U20A20218,
61525504, 61435011), the Anhui Initiative in Quantum Information
Technologies (Grant No. AHY020200), and the Youth Innovation Promotion Association
of the Chinese Academy of Sciences (Grant No. 2018490). B.-S.S acknowledges funding from the National Natural Science Foundation of China (Grant No. 11934013). We thank David MacDonald, MSc, from Liwen Bianji, Edanz Editing China (www.liwenbianji.cn/ac), for editing the English text of a draft of this manuscript.

\textbf{Author contributions. }D.-S.D. conceived the idea for the study. Z.-K.L.
implemented the physical experiments and designed the deep learning
model. Z.-K.L. wrote the manuscript. D.-S.D, B.-S.S. and, G.-C.G. supervised
the research. All authors contributed to discussions regarding the
results and analysis contained in the manuscript.

\textbf{Competing financial interests}. The authors declare no competing
financial interests.


\begin{thebibliography}{46}%
\makeatletter
\providecommand \@ifxundefined [1]{%
 \@ifx{#1\undefined}
}%
\providecommand \@ifnum [1]{%
 \ifnum #1\expandafter \@firstoftwo
 \else \expandafter \@secondoftwo
 \fi
}%
\providecommand \@ifx [1]{%
 \ifx #1\expandafter \@firstoftwo
 \else \expandafter \@secondoftwo
 \fi
}%
\providecommand \natexlab [1]{#1}%
\providecommand \enquote  [1]{``#1''}%
\providecommand \bibnamefont  [1]{#1}%
\providecommand \bibfnamefont [1]{#1}%
\providecommand \citenamefont [1]{#1}%
\providecommand \href@noop [0]{\@secondoftwo}%
\providecommand \href [0]{\begingroup \@sanitize@url \@href}%
\providecommand \@href[1]{\@@startlink{#1}\@@href}%
\providecommand \@@href[1]{\endgroup#1\@@endlink}%
\providecommand \@sanitize@url [0]{\catcode `\\12\catcode `\$12\catcode
  `\&12\catcode `\#12\catcode `\^12\catcode `\_12\catcode `\%12\relax}%
\providecommand \@@startlink[1]{}%
\providecommand \@@endlink[0]{}%
\providecommand \url  [0]{\begingroup\@sanitize@url \@url }%
\providecommand \@url [1]{\endgroup\@href {#1}{\urlprefix }}%
\providecommand \urlprefix  [0]{URL }%
\providecommand \Eprint [0]{\href }%
\providecommand \doibase [0]{http://dx.doi.org/}%
\providecommand \selectlanguage [0]{\@gobble}%
\providecommand \bibinfo  [0]{\@secondoftwo}%
\providecommand \bibfield  [0]{\@secondoftwo}%
\providecommand \translation [1]{[#1]}%
\providecommand \BibitemOpen [0]{}%
\providecommand \bibitemStop [0]{}%
\providecommand \bibitemNoStop [0]{.\EOS\space}%
\providecommand \EOS [0]{\spacefactor3000\relax}%
\providecommand \BibitemShut  [1]{\csname bibitem#1\endcsname}%
\let\auto@bib@innerbib\@empty
\bibitem [{\citenamefont {Liao}\ \emph {et~al.}(2020)\citenamefont {Liao},
  \citenamefont {Tu}, \citenamefont {Yang}, \citenamefont {Chen}, \citenamefont
  {Liu}, \citenamefont {Liang}, \citenamefont {Zhang}, \citenamefont {Yan},\
  and\ \citenamefont {Zhu}}]{EIA2020}%
  \BibitemOpen
  \bibfield  {author} {\bibinfo {author} {\bibfnamefont {Kai-Yu}\ \bibnamefont
  {Liao}}, \bibinfo {author} {\bibfnamefont {Hai-Tao}\ \bibnamefont {Tu}},
  \bibinfo {author} {\bibfnamefont {Shu-Zhe}\ \bibnamefont {Yang}}, \bibinfo
  {author} {\bibfnamefont {Chang-Jun}\ \bibnamefont {Chen}}, \bibinfo {author}
  {\bibfnamefont {Xiao-Hong}\ \bibnamefont {Liu}}, \bibinfo {author}
  {\bibfnamefont {Jie}\ \bibnamefont {Liang}}, \bibinfo {author} {\bibfnamefont
  {Xin-Ding}\ \bibnamefont {Zhang}}, \bibinfo {author} {\bibfnamefont {Hui}\
  \bibnamefont {Yan}}, \ and\ \bibinfo {author} {\bibfnamefont {Shi-Liang}\
  \bibnamefont {Zhu}},\ }\bibfield  {title} {\enquote {\bibinfo {title}
  {Microwave electrometry via electromagnetically induced absorption in cold
  {Rydberg} atoms},}\ }\href {\doibase 10.1103/PhysRevA.101.053432} {\bibfield
  {journal} {\bibinfo  {journal} {Phys. Rev. A}\ }\textbf {\bibinfo {volume}
  {101}},\ \bibinfo {pages} {053432} (\bibinfo {year} {2020})}\BibitemShut
  {NoStop}%
\bibitem [{\citenamefont {Fleischhauer}\ \emph {et~al.}(2005)\citenamefont
  {Fleischhauer}, \citenamefont {Imamoglu},\ and\ \citenamefont
  {Marangos}}]{RevModPhys.77.633}%
  \BibitemOpen
  \bibfield  {author} {\bibinfo {author} {\bibfnamefont {Michael}\ \bibnamefont
  {Fleischhauer}}, \bibinfo {author} {\bibfnamefont {Atac}\ \bibnamefont
  {Imamoglu}}, \ and\ \bibinfo {author} {\bibfnamefont {Jonathan~P.}\
  \bibnamefont {Marangos}},\ }\bibfield  {title} {\enquote {\bibinfo {title}
  {Electromagnetically induced transparency: Optics in coherent media},}\
  }\href {\doibase 10.1103/RevModPhys.77.633} {\bibfield  {journal} {\bibinfo
  {journal} {Rev. Mod. Phys.}\ }\textbf {\bibinfo {volume} {77}},\ \bibinfo
  {pages} {633--673} (\bibinfo {year} {2005})}\BibitemShut {NoStop}%
\bibitem [{\citenamefont {Holloway}\ \emph {et~al.}(2017)\citenamefont
  {Holloway}, \citenamefont {Simons}, \citenamefont {Gordon}, \citenamefont
  {Dienstfrey}, \citenamefont {Anderson},\ and\ \citenamefont
  {Raithel}}]{AT2017}%
  \BibitemOpen
  \bibfield  {author} {\bibinfo {author} {\bibfnamefont {Christopher~L.}\
  \bibnamefont {Holloway}}, \bibinfo {author} {\bibfnamefont {Matt~T.}\
  \bibnamefont {Simons}}, \bibinfo {author} {\bibfnamefont {Joshua~A.}\
  \bibnamefont {Gordon}}, \bibinfo {author} {\bibfnamefont {Andrew}\
  \bibnamefont {Dienstfrey}}, \bibinfo {author} {\bibfnamefont {David~A.}\
  \bibnamefont {Anderson}}, \ and\ \bibinfo {author} {\bibfnamefont {Georg}\
  \bibnamefont {Raithel}},\ }\bibfield  {title} {\enquote {\bibinfo {title}
  {Electric field metrology for {SI} traceability: Systematic measurement
  uncertainties in electromagnetically induced transparency in atomic vapor},}\
  }\href {\doibase 10.1063/1.4984201} {\bibfield  {journal} {\bibinfo
  {journal} {Journal of Applied Physics}\ }\textbf {\bibinfo {volume} {121}},\
  \bibinfo {pages} {233106} (\bibinfo {year} {2017})},\ \Eprint
  {http://arxiv.org/abs/https://doi.org/10.1063/1.4984201}
  {https://doi.org/10.1063/1.4984201} \BibitemShut {NoStop}%
\bibitem [{\citenamefont {Sedlacek}\ \emph {et~al.}(2012)\citenamefont
  {Sedlacek}, \citenamefont {Schwettmann}, \citenamefont {K{\"u}bler},
  \citenamefont {L{\"o}w}, \citenamefont {Pfau},\ and\ \citenamefont
  {Shaffer}}]{sedlacek2012microwave}%
  \BibitemOpen
  \bibfield  {author} {\bibinfo {author} {\bibfnamefont {Jonathon~A}\
  \bibnamefont {Sedlacek}}, \bibinfo {author} {\bibfnamefont {Arne}\
  \bibnamefont {Schwettmann}}, \bibinfo {author} {\bibfnamefont {Harald}\
  \bibnamefont {K{\"u}bler}}, \bibinfo {author} {\bibfnamefont {Robert}\
  \bibnamefont {L{\"o}w}}, \bibinfo {author} {\bibfnamefont {Tilman}\
  \bibnamefont {Pfau}}, \ and\ \bibinfo {author} {\bibfnamefont {James~P}\
  \bibnamefont {Shaffer}},\ }\bibfield  {title} {\enquote {\bibinfo {title}
  {Microwave electrometry with {Rydberg} atoms in a vapour cell using bright
  atomic resonances},}\ }\href {https://www.nature.com/articles/nphys2423}
  {\bibfield  {journal} {\bibinfo  {journal} {Nature Physics}\ }\textbf
  {\bibinfo {volume} {8}},\ \bibinfo {pages} {819--824} (\bibinfo {year}
  {2012})}\BibitemShut {NoStop}%
\bibitem [{\citenamefont {Autler}\ and\ \citenamefont {Townes}(1955)}]{AT1955}%
  \BibitemOpen
  \bibfield  {author} {\bibinfo {author} {\bibfnamefont {S.~H.}\ \bibnamefont
  {Autler}}\ and\ \bibinfo {author} {\bibfnamefont {C.~H.}\ \bibnamefont
  {Townes}},\ }\bibfield  {title} {\enquote {\bibinfo {title} {Stark effect in
  rapidly varying fields},}\ }\href {\doibase 10.1103/PhysRev.100.703}
  {\bibfield  {journal} {\bibinfo  {journal} {Phys. Rev.}\ }\textbf {\bibinfo
  {volume} {100}},\ \bibinfo {pages} {703--722} (\bibinfo {year}
  {1955})}\BibitemShut {NoStop}%
\bibitem [{\citenamefont {Abi-Salloum}(2010)}]{EITvsAT2010}%
  \BibitemOpen
  \bibfield  {author} {\bibinfo {author} {\bibfnamefont {Tony~Y.}\ \bibnamefont
  {Abi-Salloum}},\ }\bibfield  {title} {\enquote {\bibinfo {title}
  {Electromagnetically induced transparency and autler-townes splitting: Two
  similar but distinct phenomena in two categories of three-level atomic
  systems},}\ }\href {\doibase 10.1103/PhysRevA.81.053836} {\bibfield
  {journal} {\bibinfo  {journal} {Phys. Rev. A}\ }\textbf {\bibinfo {volume}
  {81}} (\bibinfo {year} {2010}),\ 10.1103/PhysRevA.81.053836}\BibitemShut
  {NoStop}%
\bibitem [{\citenamefont {Meyer}\ \emph {et~al.}(2018)\citenamefont {Meyer},
  \citenamefont {Cox}, \citenamefont {Fatemi},\ and\ \citenamefont
  {Kunz}}]{meyer2018digital}%
  \BibitemOpen
  \bibfield  {author} {\bibinfo {author} {\bibfnamefont {David~H}\ \bibnamefont
  {Meyer}}, \bibinfo {author} {\bibfnamefont {Kevin~C}\ \bibnamefont {Cox}},
  \bibinfo {author} {\bibfnamefont {Fredrik~K}\ \bibnamefont {Fatemi}}, \ and\
  \bibinfo {author} {\bibfnamefont {Paul~D}\ \bibnamefont {Kunz}},\ }\bibfield
  {title} {\enquote {\bibinfo {title} {Digital communication with {Rydberg}
  atoms and amplitude-modulated microwave fields},}\ }\href
  {https://aip.scitation.org/doi/abs/10.1063/1.5028357} {\bibfield  {journal}
  {\bibinfo  {journal} {Applied Physics Letters}\ }\textbf {\bibinfo {volume}
  {112}},\ \bibinfo {pages} {211108} (\bibinfo {year} {2018})}\BibitemShut
  {NoStop}%
\bibitem [{\citenamefont {Jiao}\ \emph {et~al.}(2019)\citenamefont {Jiao},
  \citenamefont {Han}, \citenamefont {Fan}, \citenamefont {Raithel},
  \citenamefont {Zhao},\ and\ \citenamefont {Jia}}]{jiao2019atom}%
  \BibitemOpen
  \bibfield  {author} {\bibinfo {author} {\bibfnamefont {Yuechun}\ \bibnamefont
  {Jiao}}, \bibinfo {author} {\bibfnamefont {Xiaoxuan}\ \bibnamefont {Han}},
  \bibinfo {author} {\bibfnamefont {Jiabei}\ \bibnamefont {Fan}}, \bibinfo
  {author} {\bibfnamefont {Georg}\ \bibnamefont {Raithel}}, \bibinfo {author}
  {\bibfnamefont {Jianming}\ \bibnamefont {Zhao}}, \ and\ \bibinfo {author}
  {\bibfnamefont {Suotang}\ \bibnamefont {Jia}},\ }\bibfield  {title} {\enquote
  {\bibinfo {title} {Atom-based receiver for amplitude-modulated baseband
  signals in high-frequency radio communication},}\ }\href
  {https://iopscience.iop.org/article/10.7567/1882-0786/ab5463/meta} {\bibfield
   {journal} {\bibinfo  {journal} {Applied Physics Express}\ }\textbf {\bibinfo
  {volume} {12}},\ \bibinfo {pages} {126002} (\bibinfo {year}
  {2019})}\BibitemShut {NoStop}%
\bibitem [{\citenamefont {Gordon}\ \emph {et~al.}(2019)\citenamefont {Gordon},
  \citenamefont {Simons}, \citenamefont {Haddab},\ and\ \citenamefont
  {Holloway}}]{mixer2019}%
  \BibitemOpen
  \bibfield  {author} {\bibinfo {author} {\bibfnamefont {Joshua~A.}\
  \bibnamefont {Gordon}}, \bibinfo {author} {\bibfnamefont {Matthew~T.}\
  \bibnamefont {Simons}}, \bibinfo {author} {\bibfnamefont {Abdulaziz~H.}\
  \bibnamefont {Haddab}}, \ and\ \bibinfo {author} {\bibfnamefont
  {Christopher~L.}\ \bibnamefont {Holloway}},\ }\bibfield  {title} {\enquote
  {\bibinfo {title} {Weak electric-field detection with sub-1 hz resolution at
  radio frequencies using a {Rydberg} atom-based mixer},}\ }\href {\doibase
  10.1063/1.5095633} {\bibfield  {journal} {\bibinfo  {journal} {AIP Advances}\
  }\textbf {\bibinfo {volume} {9}},\ \bibinfo {pages} {045030} (\bibinfo {year}
  {2019})},\ \Eprint {http://arxiv.org/abs/https://doi.org/10.1063/1.5095633}
  {https://doi.org/10.1063/1.5095633} \BibitemShut {NoStop}%
\bibitem [{\citenamefont {Jing}\ \emph {et~al.}(2020)\citenamefont {Jing},
  \citenamefont {Hu}, \citenamefont {Ma}, \citenamefont {Zhang}, \citenamefont
  {Zhang}, \citenamefont {Xiao},\ and\ \citenamefont {Jia}}]{jing2020atomic}%
  \BibitemOpen
  \bibfield  {author} {\bibinfo {author} {\bibfnamefont {Mingyong}\
  \bibnamefont {Jing}}, \bibinfo {author} {\bibfnamefont {Ying}\ \bibnamefont
  {Hu}}, \bibinfo {author} {\bibfnamefont {Jie}\ \bibnamefont {Ma}}, \bibinfo
  {author} {\bibfnamefont {Hao}\ \bibnamefont {Zhang}}, \bibinfo {author}
  {\bibfnamefont {Linjie}\ \bibnamefont {Zhang}}, \bibinfo {author}
  {\bibfnamefont {Liantuan}\ \bibnamefont {Xiao}}, \ and\ \bibinfo {author}
  {\bibfnamefont {Suotang}\ \bibnamefont {Jia}},\ }\bibfield  {title} {\enquote
  {\bibinfo {title} {Atomic superheterodyne receiver based on microwave-dressed
  {Rydberg} spectroscopy},}\ }\href
  {https://www.nature.com/articles/s41567-020-0918-5} {\bibfield  {journal}
  {\bibinfo  {journal} {Nature Physics}\ ,\ \bibinfo {pages} {1--5}} (\bibinfo
  {year} {2020})}\BibitemShut {NoStop}%
\bibitem [{\citenamefont {Simons}\ \emph {et~al.}(2019)\citenamefont {Simons},
  \citenamefont {Haddab}, \citenamefont {Gordon},\ and\ \citenamefont
  {Holloway}}]{simons2019rydberg}%
  \BibitemOpen
  \bibfield  {author} {\bibinfo {author} {\bibfnamefont {Matthew~T}\
  \bibnamefont {Simons}}, \bibinfo {author} {\bibfnamefont {Abdulaziz~H}\
  \bibnamefont {Haddab}}, \bibinfo {author} {\bibfnamefont {Joshua~A}\
  \bibnamefont {Gordon}}, \ and\ \bibinfo {author} {\bibfnamefont
  {Christopher~L}\ \bibnamefont {Holloway}},\ }\bibfield  {title} {\enquote
  {\bibinfo {title} {A {Rydberg} atom-based mixer: Measuring the phase of a
  radio frequency wave},}\ }\href
  {https://aip.scitation.org/doi/abs/10.1063/1.5088821} {\bibfield  {journal}
  {\bibinfo  {journal} {Applied Physics Letters}\ }\textbf {\bibinfo {volume}
  {114}},\ \bibinfo {pages} {114101} (\bibinfo {year} {2019})}\BibitemShut
  {NoStop}%
\bibitem [{\citenamefont {Holloway}\ \emph
  {et~al.}(2019{\natexlab{a}})\citenamefont {Holloway}, \citenamefont {Simons},
  \citenamefont {Gordon},\ and\ \citenamefont
  {Novotny}}]{holloway2019detecting}%
  \BibitemOpen
  \bibfield  {author} {\bibinfo {author} {\bibfnamefont {Christopher~L}\
  \bibnamefont {Holloway}}, \bibinfo {author} {\bibfnamefont {Matthew~T}\
  \bibnamefont {Simons}}, \bibinfo {author} {\bibfnamefont {Joshua~A}\
  \bibnamefont {Gordon}}, \ and\ \bibinfo {author} {\bibfnamefont {David}\
  \bibnamefont {Novotny}},\ }\bibfield  {title} {\enquote {\bibinfo {title}
  {Detecting and receiving phase-modulated signals with a {Rydberg} atom-based
  receiver},}\ }\href {\doibase 10.1109/LAWP.2019.2931450} {\bibfield
  {journal} {\bibinfo  {journal} {IEEE Antennas and Wireless Propagation
  Letters}\ }\textbf {\bibinfo {volume} {18}},\ \bibinfo {pages} {1853--1857}
  (\bibinfo {year} {2019}{\natexlab{a}})}\BibitemShut {NoStop}%
\bibitem [{\citenamefont {Holloway}\ \emph
  {et~al.}(2019{\natexlab{b}})\citenamefont {Holloway}, \citenamefont {Simons},
  \citenamefont {Haddab}, \citenamefont {Williams},\ and\ \citenamefont
  {Holloway}}]{Holloway2019guitar}%
  \BibitemOpen
  \bibfield  {author} {\bibinfo {author} {\bibfnamefont {Christopher~L.}\
  \bibnamefont {Holloway}}, \bibinfo {author} {\bibfnamefont {Matthew~T.}\
  \bibnamefont {Simons}}, \bibinfo {author} {\bibfnamefont {Abdulaziz~H.}\
  \bibnamefont {Haddab}}, \bibinfo {author} {\bibfnamefont {Carl~J.}\
  \bibnamefont {Williams}}, \ and\ \bibinfo {author} {\bibfnamefont
  {Maxwell~W.}\ \bibnamefont {Holloway}},\ }\bibfield  {title} {\enquote
  {\bibinfo {title} {A real-time guitar recording using {Rydberg} atoms and
  electromagnetically induced transparency: Quantum physics meets music},}\
  }\href {\doibase 10.1063/1.5099036} {\bibfield  {journal} {\bibinfo
  {journal} {AIP Advances}\ }\textbf {\bibinfo {volume} {9}},\ \bibinfo {pages}
  {065110} (\bibinfo {year} {2019}{\natexlab{b}})}\BibitemShut {NoStop}%
\bibitem [{\citenamefont {Robinson}\ \emph {et~al.}(2021)\citenamefont
  {Robinson}, \citenamefont {Prajapati}, \citenamefont {Senic}, \citenamefont
  {Simons},\ and\ \citenamefont {Holloway}}]{angle2021}%
  \BibitemOpen
  \bibfield  {author} {\bibinfo {author} {\bibfnamefont {Amy~K.}\ \bibnamefont
  {Robinson}}, \bibinfo {author} {\bibfnamefont {Nikunjkumar}\ \bibnamefont
  {Prajapati}}, \bibinfo {author} {\bibfnamefont {Damir}\ \bibnamefont
  {Senic}}, \bibinfo {author} {\bibfnamefont {Matthew~T.}\ \bibnamefont
  {Simons}}, \ and\ \bibinfo {author} {\bibfnamefont {Christopher~L.}\
  \bibnamefont {Holloway}},\ }\bibfield  {title} {\enquote {\bibinfo {title}
  {Determining the angle-of-arrival of a radio-frequency source with a
  {Rydberg} atom-based sensor},}\ }\href {\doibase 10.1063/5.0045601}
  {\bibfield  {journal} {\bibinfo  {journal} {Applied Physics Letters}\
  }\textbf {\bibinfo {volume} {118}},\ \bibinfo {pages} {114001} (\bibinfo
  {year} {2021})},\ \Eprint
  {http://arxiv.org/abs/https://doi.org/10.1063/5.0045601}
  {https://doi.org/10.1063/5.0045601} \BibitemShut {NoStop}%
\bibitem [{\citenamefont {Bason}\ \emph {et~al.}(2010)\citenamefont {Bason},
  \citenamefont {Tanasittikosol}, \citenamefont {Sargsyan}, \citenamefont
  {Mohapatra}, \citenamefont {Sarkisyan}, \citenamefont {Potvliege},\ and\
  \citenamefont {Adams}}]{bason2010enhanced}%
  \BibitemOpen
  \bibfield  {author} {\bibinfo {author} {\bibfnamefont {M.~G.}\ \bibnamefont
  {Bason}}, \bibinfo {author} {\bibfnamefont {M.}~\bibnamefont
  {Tanasittikosol}}, \bibinfo {author} {\bibfnamefont {A.}~\bibnamefont
  {Sargsyan}}, \bibinfo {author} {\bibfnamefont {A.~K.}\ \bibnamefont
  {Mohapatra}}, \bibinfo {author} {\bibfnamefont {D.}~\bibnamefont
  {Sarkisyan}}, \bibinfo {author} {\bibfnamefont {R.~M.}\ \bibnamefont
  {Potvliege}}, \ and\ \bibinfo {author} {\bibfnamefont {C.~S.}\ \bibnamefont
  {Adams}},\ }\bibfield  {title} {\enquote {\bibinfo {title} {Enhanced electric
  field sensitivity of rf-dressed {Rydberg} dark states},}\ }\href {\doibase
  10.1088/1367-2630/12/6/065015} {\bibfield  {journal} {\bibinfo  {journal}
  {New Journal of Physics}\ }\textbf {\bibinfo {volume} {12}} (\bibinfo {year}
  {2010}),\ 10.1088/1367-2630/12/6/065015}\BibitemShut {NoStop}%
\bibitem [{\citenamefont {Zou}\ \emph {et~al.}(2020)\citenamefont {Zou},
  \citenamefont {Song}, \citenamefont {Mu}, \citenamefont {Feng}, \citenamefont
  {Qu},\ and\ \citenamefont {Wang}}]{Multiple2020}%
  \BibitemOpen
  \bibfield  {author} {\bibinfo {author} {\bibfnamefont {Haiyang}\ \bibnamefont
  {Zou}}, \bibinfo {author} {\bibfnamefont {Zhenfei}\ \bibnamefont {Song}},
  \bibinfo {author} {\bibfnamefont {Huihui}\ \bibnamefont {Mu}}, \bibinfo
  {author} {\bibfnamefont {Zhigang}\ \bibnamefont {Feng}}, \bibinfo {author}
  {\bibfnamefont {Jifeng}\ \bibnamefont {Qu}}, \ and\ \bibinfo {author}
  {\bibfnamefont {Qilong}\ \bibnamefont {Wang}},\ }\bibfield  {title} {\enquote
  {\bibinfo {title} {Atomic receiver by utilizing multiple radio-frequency
  coupling at {Rydberg} states of rubidium},}\ }\href {\doibase
  10.3390/app10041346} {\bibfield  {journal} {\bibinfo  {journal} {Applied
  Sciences}\ }\textbf {\bibinfo {volume} {10}} (\bibinfo {year} {2020}),\
  10.3390/app10041346}\BibitemShut {NoStop}%
\bibitem [{\citenamefont {Song}\ \emph {et~al.}(2019)\citenamefont {Song},
  \citenamefont {Liu}, \citenamefont {Liu}, \citenamefont {Zhang},
  \citenamefont {Zou}, \citenamefont {Zhang},\ and\ \citenamefont
  {Qu}}]{continuously2019}%
  \BibitemOpen
  \bibfield  {author} {\bibinfo {author} {\bibfnamefont {Zhenfei}\ \bibnamefont
  {Song}}, \bibinfo {author} {\bibfnamefont {Hongping}\ \bibnamefont {Liu}},
  \bibinfo {author} {\bibfnamefont {Xiaochi}\ \bibnamefont {Liu}}, \bibinfo
  {author} {\bibfnamefont {Wanfeng}\ \bibnamefont {Zhang}}, \bibinfo {author}
  {\bibfnamefont {Haiyang}\ \bibnamefont {Zou}}, \bibinfo {author}
  {\bibfnamefont {Jie}\ \bibnamefont {Zhang}}, \ and\ \bibinfo {author}
  {\bibfnamefont {Jifeng}\ \bibnamefont {Qu}},\ }\bibfield  {title} {\enquote
  {\bibinfo {title} {Rydberg-atom-based digital communication using a
  continuously tunable radio-frequency carrier},}\ }\href {\doibase
  10.1364/OE.27.008848} {\bibfield  {journal} {\bibinfo  {journal} {Opt.
  Express}\ }\textbf {\bibinfo {volume} {27}},\ \bibinfo {pages} {8848--8857}
  (\bibinfo {year} {2019})}\BibitemShut {NoStop}%
\bibitem [{\citenamefont {Orazbayev}\ and\ \citenamefont
  {Fleury}(2020)}]{PhysRevX.10.031029}%
  \BibitemOpen
  \bibfield  {author} {\bibinfo {author} {\bibfnamefont {Bakhtiyar}\
  \bibnamefont {Orazbayev}}\ and\ \bibinfo {author} {\bibfnamefont {Romain}\
  \bibnamefont {Fleury}},\ }\bibfield  {title} {\enquote {\bibinfo {title}
  {Far-field subwavelength acoustic imaging by deep learning},}\ }\href
  {\doibase 10.1103/PhysRevX.10.031029} {\bibfield  {journal} {\bibinfo
  {journal} {Phys. Rev. X}\ }\textbf {\bibinfo {volume} {10}},\ \bibinfo
  {pages} {031029} (\bibinfo {year} {2020})}\BibitemShut {NoStop}%
\bibitem [{\citenamefont {Khanahmadi}\ and\ \citenamefont
  {M\o{}lmer}(2021)}]{khanahmadi2020timedependent}%
  \BibitemOpen
  \bibfield  {author} {\bibinfo {author} {\bibfnamefont {Maryam}\ \bibnamefont
  {Khanahmadi}}\ and\ \bibinfo {author} {\bibfnamefont {Klaus}\ \bibnamefont
  {M\o{}lmer}},\ }\bibfield  {title} {\enquote {\bibinfo {title}
  {Time-dependent atomic magnetometry with a recurrent neural network},}\
  }\href {\doibase 10.1103/PhysRevA.103.032406} {\bibfield  {journal} {\bibinfo
   {journal} {Phys. Rev. A}\ }\textbf {\bibinfo {volume} {103}},\ \bibinfo
  {pages} {032406} (\bibinfo {year} {2021})}\BibitemShut {NoStop}%
\bibitem [{\citenamefont {Giordani}\ \emph {et~al.}(2020)\citenamefont
  {Giordani}, \citenamefont {Suprano}, \citenamefont {Polino}, \citenamefont
  {Acanfora}, \citenamefont {Innocenti}, \citenamefont {Ferraro}, \citenamefont
  {Paternostro}, \citenamefont {Spagnolo},\ and\ \citenamefont
  {Sciarrino}}]{PhysRevLett.124.160401}%
  \BibitemOpen
  \bibfield  {author} {\bibinfo {author} {\bibfnamefont {Taira}\ \bibnamefont
  {Giordani}}, \bibinfo {author} {\bibfnamefont {Alessia}\ \bibnamefont
  {Suprano}}, \bibinfo {author} {\bibfnamefont {Emanuele}\ \bibnamefont
  {Polino}}, \bibinfo {author} {\bibfnamefont {Francesca}\ \bibnamefont
  {Acanfora}}, \bibinfo {author} {\bibfnamefont {Luca}\ \bibnamefont
  {Innocenti}}, \bibinfo {author} {\bibfnamefont {Alessandro}\ \bibnamefont
  {Ferraro}}, \bibinfo {author} {\bibfnamefont {Mauro}\ \bibnamefont
  {Paternostro}}, \bibinfo {author} {\bibfnamefont {Nicol\`o}\ \bibnamefont
  {Spagnolo}}, \ and\ \bibinfo {author} {\bibfnamefont {Fabio}\ \bibnamefont
  {Sciarrino}},\ }\bibfield  {title} {\enquote {\bibinfo {title} {Machine
  learning-based classification of vector vortex beams},}\ }\href {\doibase
  10.1103/PhysRevLett.124.160401} {\bibfield  {journal} {\bibinfo  {journal}
  {Phys. Rev. Lett.}\ }\textbf {\bibinfo {volume} {124}},\ \bibinfo {pages}
  {160401} (\bibinfo {year} {2020})}\BibitemShut {NoStop}%
\bibitem [{\citenamefont {Liu}\ \emph {et~al.}(2019)\citenamefont {Liu},
  \citenamefont {Yan}, \citenamefont {Liu},\ and\ \citenamefont
  {Chen}}]{PhysRevLett.123.183902}%
  \BibitemOpen
  \bibfield  {author} {\bibinfo {author} {\bibfnamefont {Zhanwei}\ \bibnamefont
  {Liu}}, \bibinfo {author} {\bibfnamefont {Shuo}\ \bibnamefont {Yan}},
  \bibinfo {author} {\bibfnamefont {Haigang}\ \bibnamefont {Liu}}, \ and\
  \bibinfo {author} {\bibfnamefont {Xianfeng}\ \bibnamefont {Chen}},\
  }\bibfield  {title} {\enquote {\bibinfo {title} {Superhigh-resolution
  recognition of optical vortex modes assisted by a deep-learning method},}\
  }\href {\doibase 10.1103/PhysRevLett.123.183902} {\bibfield  {journal}
  {\bibinfo  {journal} {Phys. Rev. Lett.}\ }\textbf {\bibinfo {volume} {123}},\
  \bibinfo {pages} {183902} (\bibinfo {year} {2019})}\BibitemShut {NoStop}%
\bibitem [{\citenamefont {Doster}\ and\ \citenamefont
  {Watnik}(2017)}]{ISI:000399872100026}%
  \BibitemOpen
  \bibfield  {author} {\bibinfo {author} {\bibfnamefont {Timothy}\ \bibnamefont
  {Doster}}\ and\ \bibinfo {author} {\bibfnamefont {Abbie~T.}\ \bibnamefont
  {Watnik}},\ }\bibfield  {title} {\enquote {\bibinfo {title} {Machine learning
  approach to oam beam demultiplexing via convolutional neural networks},}\
  }\href {\doibase 10.1364/AO.56.003386} {\bibfield  {journal} {\bibinfo
  {journal} {Appl. Opt.}\ }\textbf {\bibinfo {volume} {56}},\ \bibinfo {pages}
  {3386--3396} (\bibinfo {year} {2017})}\BibitemShut {NoStop}%
\bibitem [{\citenamefont {da~Silva}\ \emph {et~al.}(2021)\citenamefont
  {da~Silva}, \citenamefont {Marques}, \citenamefont {Rodrigues}, \citenamefont
  {Ribeiro},\ and\ \citenamefont {Khoury}}]{MLOAM}%
  \BibitemOpen
  \bibfield  {author} {\bibinfo {author} {\bibfnamefont {B.~Pinheiro}\
  \bibnamefont {da~Silva}}, \bibinfo {author} {\bibfnamefont {B.~A.~D.}\
  \bibnamefont {Marques}}, \bibinfo {author} {\bibfnamefont {R.~B.}\
  \bibnamefont {Rodrigues}}, \bibinfo {author} {\bibfnamefont {P.~H.~Souto}\
  \bibnamefont {Ribeiro}}, \ and\ \bibinfo {author} {\bibfnamefont {A.~Z.}\
  \bibnamefont {Khoury}},\ }\bibfield  {title} {\enquote {\bibinfo {title}
  {Machine-learning recognition of light orbital-angular-momentum
  superpositions},}\ }\href {\doibase 10.1103/PhysRevA.103.063704} {\bibfield
  {journal} {\bibinfo  {journal} {Phys. Rev. A}\ }\textbf {\bibinfo {volume}
  {103}},\ \bibinfo {pages} {063704} (\bibinfo {year} {2021})}\BibitemShut
  {NoStop}%
\bibitem [{\citenamefont {Wigley}\ \emph {et~al.}(2016)\citenamefont {Wigley},
  \citenamefont {Everitt}, \citenamefont {van~den Hengel}, \citenamefont
  {Bastian}, \citenamefont {Sooriyabandara}, \citenamefont {McDonald},
  \citenamefont {Hardman}, \citenamefont {Quinlivan}, \citenamefont {Manju},
  \citenamefont {Kuhn},\ and\ \citenamefont {et~al.}}]{Wigley_2016}%
  \BibitemOpen
  \bibfield  {author} {\bibinfo {author} {\bibfnamefont {P.~B.}\ \bibnamefont
  {Wigley}}, \bibinfo {author} {\bibfnamefont {P.~J.}\ \bibnamefont {Everitt}},
  \bibinfo {author} {\bibfnamefont {A.}~\bibnamefont {van~den Hengel}},
  \bibinfo {author} {\bibfnamefont {J.~W.}\ \bibnamefont {Bastian}}, \bibinfo
  {author} {\bibfnamefont {M.~A.}\ \bibnamefont {Sooriyabandara}}, \bibinfo
  {author} {\bibfnamefont {G.~D.}\ \bibnamefont {McDonald}}, \bibinfo {author}
  {\bibfnamefont {K.~S.}\ \bibnamefont {Hardman}}, \bibinfo {author}
  {\bibfnamefont {C.~D.}\ \bibnamefont {Quinlivan}}, \bibinfo {author}
  {\bibfnamefont {P.}~\bibnamefont {Manju}}, \bibinfo {author} {\bibfnamefont
  {C.~C.~N.}\ \bibnamefont {Kuhn}}, \ and\ \bibinfo {author} {\bibnamefont
  {et~al.}},\ }\bibfield  {title} {\enquote {\bibinfo {title} {Fast
  machine-learning online optimization of ultra-cold-atom experiments},}\
  }\href {\doibase 10.1038/srep25890} {\bibfield  {journal} {\bibinfo
  {journal} {Scientific Reports}\ }\textbf {\bibinfo {volume} {6}} (\bibinfo
  {year} {2016}),\ 10.1038/srep25890}\BibitemShut {NoStop}%
\bibitem [{\citenamefont {Tranter}\ \emph {et~al.}(2018)\citenamefont
  {Tranter}, \citenamefont {Slatyer}, \citenamefont {Hush},\ and\ \citenamefont
  {et~al.}}]{Deep_mot2018}%
  \BibitemOpen
  \bibfield  {author} {\bibinfo {author} {\bibfnamefont {A.~D.}\ \bibnamefont
  {Tranter}}, \bibinfo {author} {\bibfnamefont {H.~J.}\ \bibnamefont
  {Slatyer}}, \bibinfo {author} {\bibfnamefont {M.~R.}\ \bibnamefont {Hush}}, \
  and\ \bibinfo {author} {\bibnamefont {et~al.}},\ }\bibfield  {title}
  {\enquote {\bibinfo {title} {Multiparameter optimisation of a magneto-optical
  trap using deep learning},}\ }\href {\doibase 10.1038/s41467-018-06847-1}
  {\bibfield  {journal} {\bibinfo  {journal} {Nature Communications}\ }\textbf
  {\bibinfo {volume} {9}} (\bibinfo {year} {2018}),\
  10.1038/s41467-018-06847-1}\BibitemShut {NoStop}%
\bibitem [{\citenamefont {Mukherjee}\ \emph {et~al.}(2020)\citenamefont
  {Mukherjee}, \citenamefont {Xie},\ and\ \citenamefont
  {Mintert}}]{BayesianGHZ2020}%
  \BibitemOpen
  \bibfield  {author} {\bibinfo {author} {\bibfnamefont {Rick}\ \bibnamefont
  {Mukherjee}}, \bibinfo {author} {\bibfnamefont {Harry}\ \bibnamefont {Xie}},
  \ and\ \bibinfo {author} {\bibfnamefont {Florian}\ \bibnamefont {Mintert}},\
  }\bibfield  {title} {\enquote {\bibinfo {title} {Bayesian optimal control of
  greenberger-horne-zeilinger states in {Rydberg} lattices},}\ }\href {\doibase
  10.1103/PhysRevLett.125.203603} {\bibfield  {journal} {\bibinfo  {journal}
  {Phys. Rev. Lett.}\ }\textbf {\bibinfo {volume} {125}},\ \bibinfo {pages}
  {203603} (\bibinfo {year} {2020})}\BibitemShut {NoStop}%
\bibitem [{\citenamefont {Mills}\ \emph {et~al.}(2020)\citenamefont {Mills},
  \citenamefont {Ronagh},\ and\ \citenamefont {Tamblyn}}]{Mills_2020Cool}%
  \BibitemOpen
  \bibfield  {author} {\bibinfo {author} {\bibfnamefont {Kyle}\ \bibnamefont
  {Mills}}, \bibinfo {author} {\bibfnamefont {Pooya}\ \bibnamefont {Ronagh}}, \
  and\ \bibinfo {author} {\bibfnamefont {Isaac}\ \bibnamefont {Tamblyn}},\
  }\bibfield  {title} {\enquote {\bibinfo {title} {Finding the ground state of
  spin hamiltonians with reinforcement learning},}\ }\href {\doibase
  10.1038/s42256-020-0226-x} {\bibfield  {journal} {\bibinfo  {journal} {Nature
  Machine Intelligence}\ }\textbf {\bibinfo {volume} {2}},\ \bibinfo {pages}
  {509--517} (\bibinfo {year} {2020})}\BibitemShut {NoStop}%
\bibitem [{\citenamefont {Wang}\ \emph {et~al.}(2020)\citenamefont {Wang},
  \citenamefont {Ashida},\ and\ \citenamefont {Ueda}}]{Cartpoles2020}%
  \BibitemOpen
  \bibfield  {author} {\bibinfo {author} {\bibfnamefont {Zhikang~T.}\
  \bibnamefont {Wang}}, \bibinfo {author} {\bibfnamefont {Yuto}\ \bibnamefont
  {Ashida}}, \ and\ \bibinfo {author} {\bibfnamefont {Masahito}\ \bibnamefont
  {Ueda}},\ }\bibfield  {title} {\enquote {\bibinfo {title} {Deep reinforcement
  learning control of quantum cartpoles},}\ }\href {\doibase
  10.1103/PhysRevLett.125.100401} {\bibfield  {journal} {\bibinfo  {journal}
  {Phys. Rev. Lett.}\ }\textbf {\bibinfo {volume} {125}},\ \bibinfo {pages}
  {100401} (\bibinfo {year} {2020})}\BibitemShut {NoStop}%
\bibitem [{\citenamefont {Bukov}\ \emph {et~al.}(2018)\citenamefont {Bukov},
  \citenamefont {Day}, \citenamefont {Sels}, \citenamefont {Weinberg},
  \citenamefont {Polkovnikov},\ and\ \citenamefont {Mehta}}]{Bukov_2018Rl}%
  \BibitemOpen
  \bibfield  {author} {\bibinfo {author} {\bibfnamefont {Marin}\ \bibnamefont
  {Bukov}}, \bibinfo {author} {\bibfnamefont {Alexandre G.~R.}\ \bibnamefont
  {Day}}, \bibinfo {author} {\bibfnamefont {Dries}\ \bibnamefont {Sels}},
  \bibinfo {author} {\bibfnamefont {Phillip}\ \bibnamefont {Weinberg}},
  \bibinfo {author} {\bibfnamefont {Anatoli}\ \bibnamefont {Polkovnikov}}, \
  and\ \bibinfo {author} {\bibfnamefont {Pankaj}\ \bibnamefont {Mehta}},\
  }\bibfield  {title} {\enquote {\bibinfo {title} {Reinforcement learning in
  different phases of quantum control},}\ }\href {\doibase
  10.1103/physrevx.8.031086} {\bibfield  {journal} {\bibinfo  {journal} {Phys.
  Rev. X}\ }\textbf {\bibinfo {volume} {8}} (\bibinfo {year} {2018}),\
  10.1103/physrevx.8.031086}\BibitemShut {NoStop}%
\bibitem [{\citenamefont {{\v{S}}ibali{\'c}}\ \emph {et~al.}(2017)\citenamefont
  {{\v{S}}ibali{\'c}}, \citenamefont {Pritchard}, \citenamefont {Adams},\ and\
  \citenamefont {Weatherill}}]{vsibalic2017arc}%
  \BibitemOpen
  \bibfield  {author} {\bibinfo {author} {\bibfnamefont {Nikola}\ \bibnamefont
  {{\v{S}}ibali{\'c}}}, \bibinfo {author} {\bibfnamefont {Jonathan~D}\
  \bibnamefont {Pritchard}}, \bibinfo {author} {\bibfnamefont {Charles~S}\
  \bibnamefont {Adams}}, \ and\ \bibinfo {author} {\bibfnamefont {Kevin~J}\
  \bibnamefont {Weatherill}},\ }\bibfield  {title} {\enquote {\bibinfo {title}
  {Arc: An open-source library for calculating properties of alkali {Rydberg}
  atoms},}\ }\href {\doibase http://dx.doi.org/10.17632/hm5n8w628c.} {\bibfield
   {journal} {\bibinfo  {journal} {Computer Physics Communications}\ }\textbf
  {\bibinfo {volume} {220}},\ \bibinfo {pages} {319--331} (\bibinfo {year}
  {2017})}\BibitemShut {NoStop}%
\bibitem [{\citenamefont {Chollet}\ \emph {et~al.}(2015)\citenamefont {Chollet}
  \emph {et~al.}}]{chollet2015keras}%
  \BibitemOpen
  \bibfield  {author} {\bibinfo {author} {\bibfnamefont {Fran\c{c}ois}\
  \bibnamefont {Chollet}} \emph {et~al.},\ }\href@noop {} {\enquote {\bibinfo
  {title} {Keras},}\ }\bibinfo {howpublished}
  {\url{https://github.com/fchollet/keras}} (\bibinfo {year}
  {2015})\BibitemShut {NoStop}%
\bibitem [{\citenamefont {Pedregosa}\ \emph {et~al.}(2011)\citenamefont
  {Pedregosa}, \citenamefont {Varoquaux}, \citenamefont {Gramfort},
  \citenamefont {Michel}, \citenamefont {Thirion}, \citenamefont {Grisel},
  \citenamefont {Blondel}, \citenamefont {Prettenhofer}, \citenamefont {Weiss},
  \citenamefont {Dubourg}, \citenamefont {Vanderplas}, \citenamefont {Passos},
  \citenamefont {Cournapeau}, \citenamefont {Brucher}, \citenamefont {Perrot},\
  and\ \citenamefont {Duchesnay}}]{scikit_learn}%
  \BibitemOpen
  \bibfield  {author} {\bibinfo {author} {\bibfnamefont {F.}~\bibnamefont
  {Pedregosa}}, \bibinfo {author} {\bibfnamefont {G.}~\bibnamefont
  {Varoquaux}}, \bibinfo {author} {\bibfnamefont {A.}~\bibnamefont {Gramfort}},
  \bibinfo {author} {\bibfnamefont {V.}~\bibnamefont {Michel}}, \bibinfo
  {author} {\bibfnamefont {B.}~\bibnamefont {Thirion}}, \bibinfo {author}
  {\bibfnamefont {O.}~\bibnamefont {Grisel}}, \bibinfo {author} {\bibfnamefont
  {M.}~\bibnamefont {Blondel}}, \bibinfo {author} {\bibfnamefont
  {P.}~\bibnamefont {Prettenhofer}}, \bibinfo {author} {\bibfnamefont
  {R.}~\bibnamefont {Weiss}}, \bibinfo {author} {\bibfnamefont
  {V.}~\bibnamefont {Dubourg}}, \bibinfo {author} {\bibfnamefont
  {J.}~\bibnamefont {Vanderplas}}, \bibinfo {author} {\bibfnamefont
  {A.}~\bibnamefont {Passos}}, \bibinfo {author} {\bibfnamefont
  {D.}~\bibnamefont {Cournapeau}}, \bibinfo {author} {\bibfnamefont
  {M.}~\bibnamefont {Brucher}}, \bibinfo {author} {\bibfnamefont
  {M.}~\bibnamefont {Perrot}}, \ and\ \bibinfo {author} {\bibfnamefont
  {E.}~\bibnamefont {Duchesnay}},\ }\bibfield  {title} {\enquote {\bibinfo
  {title} {Scikit-learn: Machine learning in {P}ython},}\ }\href@noop {}
  {\bibfield  {journal} {\bibinfo  {journal} {Journal of Machine Learning
  Research}\ }\textbf {\bibinfo {volume} {12}},\ \bibinfo {pages} {2825--2830}
  (\bibinfo {year} {2011})}\BibitemShut {NoStop}%
\bibitem [{\citenamefont {Chollet}(2017)}]{deep_learning_with_python}%
  \BibitemOpen
  \bibfield  {author} {\bibinfo {author} {\bibfnamefont {Francois}\
  \bibnamefont {Chollet}},\ }\href@noop {} {\emph {\bibinfo {title} {Deep
  learning with python}}}\ (\bibinfo  {publisher} {Manning Publications},\
  \bibinfo {year} {2017})\BibitemShut {NoStop}%
\bibitem [{\citenamefont {Sahoo}\ \emph {et~al.}(2017)\citenamefont {Sahoo},
  \citenamefont {Pham}, \citenamefont {Lu},\ and\ \citenamefont
  {Hoi}}]{sahoo2017online}%
  \BibitemOpen
  \bibfield  {author} {\bibinfo {author} {\bibfnamefont {Doyen}\ \bibnamefont
  {Sahoo}}, \bibinfo {author} {\bibfnamefont {Quang}\ \bibnamefont {Pham}},
  \bibinfo {author} {\bibfnamefont {Jing}\ \bibnamefont {Lu}}, \ and\ \bibinfo
  {author} {\bibfnamefont {Steven C.~H.}\ \bibnamefont {Hoi}},\ }\href@noop {}
  {\enquote {\bibinfo {title} {Online deep learning: Learning deep neural
  networks on the fly},}\ } (\bibinfo {year} {2017}),\ \Eprint
  {http://arxiv.org/abs/1711.03705} {arXiv:1711.03705 [cs.LG]} \BibitemShut
  {NoStop}%
\bibitem [{\citenamefont {Cox}\ \emph {et~al.}(2018)\citenamefont {Cox},
  \citenamefont {Meyer}, \citenamefont {Fatemi},\ and\ \citenamefont
  {Kunz}}]{cox2018quantum}%
  \BibitemOpen
  \bibfield  {author} {\bibinfo {author} {\bibfnamefont {Kevin~C}\ \bibnamefont
  {Cox}}, \bibinfo {author} {\bibfnamefont {David~H}\ \bibnamefont {Meyer}},
  \bibinfo {author} {\bibfnamefont {Fredrik~K}\ \bibnamefont {Fatemi}}, \ and\
  \bibinfo {author} {\bibfnamefont {Paul~D}\ \bibnamefont {Kunz}},\ }\bibfield
  {title} {\enquote {\bibinfo {title} {Quantum-limited atomic receiver in the
  electrically small regime},}\ }\href
  {https://journals.aps.org/prl/abstract/10.1103/PhysRevLett.121.110502}
  {\bibfield  {journal} {\bibinfo  {journal} {Phys. Rev. Lett.}\ }\textbf
  {\bibinfo {volume} {121}},\ \bibinfo {pages} {110502} (\bibinfo {year}
  {2018})}\BibitemShut {NoStop}%
\bibitem [{\citenamefont {Kumar}\ \emph {et~al.}(2017)\citenamefont {Kumar},
  \citenamefont {Fan}, \citenamefont {K\"{u}bler}, \citenamefont {Jahangiri},\
  and\ \citenamefont {Shaffer}}]{Kumar17}%
  \BibitemOpen
  \bibfield  {author} {\bibinfo {author} {\bibfnamefont {Santosh}\ \bibnamefont
  {Kumar}}, \bibinfo {author} {\bibfnamefont {Haoquan}\ \bibnamefont {Fan}},
  \bibinfo {author} {\bibfnamefont {Harald}\ \bibnamefont {K\"{u}bler}},
  \bibinfo {author} {\bibfnamefont {Akbar~J.}\ \bibnamefont {Jahangiri}}, \
  and\ \bibinfo {author} {\bibfnamefont {James~P.}\ \bibnamefont {Shaffer}},\
  }\bibfield  {title} {\enquote {\bibinfo {title} {Rydberg-atom based
  radio-frequency electrometry using frequency modulation spectroscopy in room
  temperature vapor cells},}\ }\href {\doibase 10.1364/OE.25.008625} {\bibfield
   {journal} {\bibinfo  {journal} {Opt. Express}\ }\textbf {\bibinfo {volume}
  {25}},\ \bibinfo {pages} {8625--8637} (\bibinfo {year} {2017})}\BibitemShut
  {NoStop}%
\bibitem [{\citenamefont {Meyer}\ \emph {et~al.}(2021)\citenamefont {Meyer},
  \citenamefont {Kunz},\ and\ \citenamefont {Cox}}]{PhysRevApplied.15.014053}%
  \BibitemOpen
  \bibfield  {author} {\bibinfo {author} {\bibfnamefont {David~H.}\
  \bibnamefont {Meyer}}, \bibinfo {author} {\bibfnamefont {Paul~D.}\
  \bibnamefont {Kunz}}, \ and\ \bibinfo {author} {\bibfnamefont {Kevin~C.}\
  \bibnamefont {Cox}},\ }\bibfield  {title} {\enquote {\bibinfo {title}
  {Waveguide-coupled {Rydberg} spectrum analyzer from 0 to 20 ghz},}\ }\href
  {\doibase 10.1103/PhysRevApplied.15.014053} {\bibfield  {journal} {\bibinfo
  {journal} {Phys. Rev. Applied}\ }\textbf {\bibinfo {volume} {15}},\ \bibinfo
  {pages} {014053} (\bibinfo {year} {2021})}\BibitemShut {NoStop}%
\bibitem [{\citenamefont {Meyer}\ \emph {et~al.}(2020)\citenamefont {Meyer},
  \citenamefont {Castillo}, \citenamefont {Cox},\ and\ \citenamefont
  {Kunz}}]{Meyer_2020}%
  \BibitemOpen
  \bibfield  {author} {\bibinfo {author} {\bibfnamefont {David~H}\ \bibnamefont
  {Meyer}}, \bibinfo {author} {\bibfnamefont {Zachary~A}\ \bibnamefont
  {Castillo}}, \bibinfo {author} {\bibfnamefont {Kevin~C}\ \bibnamefont {Cox}},
  \ and\ \bibinfo {author} {\bibfnamefont {Paul~D}\ \bibnamefont {Kunz}},\
  }\bibfield  {title} {\enquote {\bibinfo {title} {Assessment of {Rydberg}
  atoms for wideband electric field sensing},}\ }\href {\doibase
  10.1088/1361-6455/ab6051} {\bibfield  {journal} {\bibinfo  {journal} {Journal
  of Physics B: Atomic, Molecular and Optical Physics}\ }\textbf {\bibinfo
  {volume} {53}},\ \bibinfo {pages} {034001} (\bibinfo {year}
  {2020})}\BibitemShut {NoStop}%
\bibitem [{\citenamefont {Wade}\ \emph {et~al.}(2018)\citenamefont {Wade},
  \citenamefont {Marcuzzi}, \citenamefont {Levi}, \citenamefont {Kondo},
  \citenamefont {Lesanovsky}, \citenamefont {Adams},\ and\ \citenamefont
  {Weatherill}}]{wade2018terahertz}%
  \BibitemOpen
  \bibfield  {author} {\bibinfo {author} {\bibfnamefont {Christopher~G}\
  \bibnamefont {Wade}}, \bibinfo {author} {\bibfnamefont {M}~\bibnamefont
  {Marcuzzi}}, \bibinfo {author} {\bibfnamefont {E}~\bibnamefont {Levi}},
  \bibinfo {author} {\bibfnamefont {Jorge~M}\ \bibnamefont {Kondo}}, \bibinfo
  {author} {\bibfnamefont {Igor}\ \bibnamefont {Lesanovsky}}, \bibinfo {author}
  {\bibfnamefont {Charles~S}\ \bibnamefont {Adams}}, \ and\ \bibinfo {author}
  {\bibfnamefont {Kevin~J}\ \bibnamefont {Weatherill}},\ }\bibfield  {title}
  {\enquote {\bibinfo {title} {A terahertz-driven non-equilibrium phase
  transition in a room temperature atomic vapour},}\ }\href
  {https://doi.org/10.15128/r2s1784k74g} {\bibfield  {journal} {\bibinfo
  {journal} {Nature Communications}\ }\textbf {\bibinfo {volume} {9}},\
  \bibinfo {pages} {3567} (\bibinfo {year} {2018})}\BibitemShut {NoStop}%
\bibitem [{\citenamefont {Holloway}\ \emph {et~al.}(2014)\citenamefont
  {Holloway}, \citenamefont {Gordon}, \citenamefont {Jefferts}, \citenamefont
  {Schwarzkopf}, \citenamefont {Anderson}, \citenamefont {Miller},
  \citenamefont {Thaicharoen},\ and\ \citenamefont
  {Raithel}}]{holloway2014broadband}%
  \BibitemOpen
  \bibfield  {author} {\bibinfo {author} {\bibfnamefont {Christopher~L.}\
  \bibnamefont {Holloway}}, \bibinfo {author} {\bibfnamefont {Joshua~A.}\
  \bibnamefont {Gordon}}, \bibinfo {author} {\bibfnamefont {Steven}\
  \bibnamefont {Jefferts}}, \bibinfo {author} {\bibfnamefont {Andrew}\
  \bibnamefont {Schwarzkopf}}, \bibinfo {author} {\bibfnamefont {David~A.}\
  \bibnamefont {Anderson}}, \bibinfo {author} {\bibfnamefont {Stephanie~A.}\
  \bibnamefont {Miller}}, \bibinfo {author} {\bibfnamefont {Nithiwadee}\
  \bibnamefont {Thaicharoen}}, \ and\ \bibinfo {author} {\bibfnamefont {Georg}\
  \bibnamefont {Raithel}},\ }\bibfield  {title} {\enquote {\bibinfo {title}
  {Broadband {R}ydberg atom-based electric-field probe for {SI}-traceable,
  self-calibrated measurements},}\ }\href {\doibase 10.1109/TAP.2014.2360208}
  {\bibfield  {journal} {\bibinfo  {journal} {IEEE Transactions on Antennas and
  Propagation}\ }\textbf {\bibinfo {volume} {62}},\ \bibinfo {pages}
  {6169--6182} (\bibinfo {year} {2014})}\BibitemShut {NoStop}%
\bibitem [{\citenamefont {Ioffe}\ and\ \citenamefont {Szegedy}(2015)}]{Batch}%
  \BibitemOpen
  \bibfield  {author} {\bibinfo {author} {\bibfnamefont {Sergey}\ \bibnamefont
  {Ioffe}}\ and\ \bibinfo {author} {\bibfnamefont {Christian}\ \bibnamefont
  {Szegedy}},\ }\bibfield  {title} {\enquote {\bibinfo {title} {Batch
  normalization: Accelerating deep network training by reducing internal
  covariate shift},}\ }in\ \href
  {http://proceedings.mlr.press/v37/ioffe15.html} {\emph {\bibinfo {booktitle}
  {Proceedings of the 32nd International Conference on Machine Learning}}},\
  \bibinfo {series} {Proceedings of Machine Learning Research}, Vol.~\bibinfo
  {volume} {37},\ \bibinfo {editor} {edited by\ \bibinfo {editor}
  {\bibfnamefont {Francis}\ \bibnamefont {Bach}}\ and\ \bibinfo {editor}
  {\bibfnamefont {David}\ \bibnamefont {Blei}}}\ (\bibinfo  {publisher}
  {PMLR},\ \bibinfo {address} {Lille, France},\ \bibinfo {year} {2015})\ pp.\
  \bibinfo {pages} {448--456}\BibitemShut {NoStop}%
\bibitem [{\citenamefont {Hochreiter}\ and\ \citenamefont
  {Schmidhuber}(1997)}]{LSTM}%
  \BibitemOpen
  \bibfield  {author} {\bibinfo {author} {\bibfnamefont {S.}~\bibnamefont
  {Hochreiter}}\ and\ \bibinfo {author} {\bibfnamefont {J.}~\bibnamefont
  {Schmidhuber}},\ }\bibfield  {title} {\enquote {\bibinfo {title} {{Long
  Short-Term Memory}},}\ }\href {\doibase 10.1162/neco.1997.9.8.1735}
  {\bibfield  {journal} {\bibinfo  {journal} {Neural Computation}\ }\textbf
  {\bibinfo {volume} {9}},\ \bibinfo {pages} {1735--1780} (\bibinfo {year}
  {1997})},\ \Eprint
  {http://arxiv.org/abs/https://direct.mit.edu/neco/article-pdf/9/8/1735/813796/neco.1997.9.8.1735.pdf}
  {https://direct.mit.edu/neco/article-pdf/9/8/1735/813796/neco.1997.9.8.1735.pdf}
  \BibitemShut {NoStop}%
\bibitem [{\citenamefont {Geoffrey}\ \emph {et~al.}(2012)\citenamefont
  {Geoffrey}, \citenamefont {Nitish},\ and\ \citenamefont
  {Kevin}}]{mini_batch}%
  \BibitemOpen
  \bibfield  {author} {\bibinfo {author} {\bibfnamefont {Hinton}\ \bibnamefont
  {Geoffrey}}, \bibinfo {author} {\bibfnamefont {Srivastava}\ \bibnamefont
  {Nitish}}, \ and\ \bibinfo {author} {\bibfnamefont {Swersky}\ \bibnamefont
  {Kevin}},\ }\href@noop {} {\enquote {\bibinfo {title} {Lecture 6a overview of
  mini-batch gradient descent},}\ }\bibinfo {howpublished}
  {\url{http://www.cs.toronto.edu/~tijmen/csc321/slides/lecture_slides_lec6.pdf}}
  (\bibinfo {year} {2012})\BibitemShut {NoStop}%
\bibitem [{\citenamefont {Akiba}\ \emph {et~al.}(2019)\citenamefont {Akiba},
  \citenamefont {Sano}, \citenamefont {Yanase}, \citenamefont {Ohta},\ and\
  \citenamefont {Koyama}}]{optuna_2019}%
  \BibitemOpen
  \bibfield  {author} {\bibinfo {author} {\bibfnamefont {Takuya}\ \bibnamefont
  {Akiba}}, \bibinfo {author} {\bibfnamefont {Shotaro}\ \bibnamefont {Sano}},
  \bibinfo {author} {\bibfnamefont {Toshihiko}\ \bibnamefont {Yanase}},
  \bibinfo {author} {\bibfnamefont {Takeru}\ \bibnamefont {Ohta}}, \ and\
  \bibinfo {author} {\bibfnamefont {Masanori}\ \bibnamefont {Koyama}},\
  }\bibfield  {title} {\enquote {\bibinfo {title} {Optuna: A next-generation
  hyperparameter optimization framework},}\ }in\ \href@noop {} {\emph {\bibinfo
  {booktitle} {Proceedings of the 25rd {ACM} {SIGKDD} International Conference
  on Knowledge Discovery and Data Mining}}}\ (\bibinfo {year}
  {2019})\BibitemShut {NoStop}%
\bibitem [{\citenamefont {Vaswani}\ \emph {et~al.}(2017)\citenamefont
  {Vaswani}, \citenamefont {Shazeer}, \citenamefont {Parmar}, \citenamefont
  {Uszkoreit}, \citenamefont {Jones}, \citenamefont {Gomez}, \citenamefont
  {Kaiser},\ and\ \citenamefont {Polosukhin}}]{vaswani2017attention}%
  \BibitemOpen
  \bibfield  {author} {\bibinfo {author} {\bibfnamefont {Ashish}\ \bibnamefont
  {Vaswani}}, \bibinfo {author} {\bibfnamefont {Noam}\ \bibnamefont {Shazeer}},
  \bibinfo {author} {\bibfnamefont {Niki}\ \bibnamefont {Parmar}}, \bibinfo
  {author} {\bibfnamefont {Jakob}\ \bibnamefont {Uszkoreit}}, \bibinfo {author}
  {\bibfnamefont {Llion}\ \bibnamefont {Jones}}, \bibinfo {author}
  {\bibfnamefont {Aidan~N.}\ \bibnamefont {Gomez}}, \bibinfo {author}
  {\bibfnamefont {Lukasz}\ \bibnamefont {Kaiser}}, \ and\ \bibinfo {author}
  {\bibfnamefont {Illia}\ \bibnamefont {Polosukhin}},\ }\href@noop {} {\enquote
  {\bibinfo {title} {Attention is all you need},}\ } (\bibinfo {year} {2017}),\
  \Eprint {http://arxiv.org/abs/1706.03762} {arXiv:1706.03762 [cs.CL]}
  \BibitemShut {NoStop}%
\bibitem [{\citenamefont {Liu}(2022)}]{Code}%
  \BibitemOpen
  \bibfield  {author} {\bibinfo {author} {\bibfnamefont {Zong-Kai}\
  \bibnamefont {Liu}},\ }\href@noop {} {\enquote {\bibinfo {title} {Deep
  learning enhanced {Rydberg} multifrequency microwave recognition},}\
  }\bibinfo {howpublished} {\url{ https://doi.org/10.5281/zenodo.6202552 }}
  (\bibinfo {year} {2022})\BibitemShut {NoStop}%
\end{thebibliography}
\end{document}


\begin{centering}
\textbf{Supplementary materials for: Deep learning enhanced Rydberg multifrequency microwave recognition}

\end{centering}

\vspace{3em}

\textbf{Noise analysis. }The systematic noise spectra are shown in Fig.~\ref{spectrum}.
The systematic noise comes from two sides: the first type is from the exterior, such as 1/f
noise of electric circuits, background electric noise, and the noise from the
mechanical vibration, etc.; the second type is from the interior (from atoms in the vapor cell)
such as the transit noise due to thermal atoms. As stated in the literature~\citep{jing2020atomic}, there are different types of noise for different frequencies.  For frequency below 1 kHz, there
is 1/f noise. Then, for a higher frequency range (1 kHz$\sim$100 kHz), there is noise from atomic transitions out from and into the light area. At much higher frequencies (100 kHz$\sim$1 MHz), the noise comes from the control systems of the lasers. During the experiments, we found that our deep learning model was robust with respect to these noise types. In fact, the deep learning model extracts the signal from noisy channel with high accuracy regardless of where the noise come from.

Under the systematic noise and the additional noise, the transmission spectrum are shown in Fig.~\ref{spectrum_scale}, where the additional noise is increased by adjusting its standard deviation $\sigma$. The fitting results of the master equation are also shown, which demonstrate that for larger noise the fitting curve of the master equation is distinct from the signal.

\textbf{Visualization the intermediate results of the deep learning model.} After passing through the max-pooling layer, the data treated in the deep learning model are visualized directly, as shown in Fig. 7 of the main text. The data are expanded by the one-dimensional convolution kernel from one dimension to 20 dimensions. These new dimensions act as new features of the data. The feature space is then provided after the features of the intermediate results are reduced through principal component analysis (PCA) and T-distributed Stochastic Neighbor Embedding (t-SNE) ~\citep{scikit_learn}. PCA is a method of linearly reducing the dimensionality to project data to a lower dimensional space whereas t-SNE is a method of nonlinearly reducing the dimensionality. The feature space is obtained after the intermediate results are fed into these new models (even a linear method such as PCA). Signals carrying the same message are inclined to be together whereas those carrying different messages are separated, as shown in Fig.~\ref{pca}. The features captured by the deep learning model are thus transferable and compatible with other models. Figure~\ref{pca} also presents dimensionality reduction results for noisy data, showing that the capturing of features by the deep learning model is robust against additional noise.

We then refer to the Grad-CAM method, which is popular in the field of computer vision and used to highlight where the focus of a model is on an image~\citep{2019Grad}. Heat maps are presented in Fig.~\ref{heatmap}, where the transmission spectrum and the model¡¯s focus are given. The area by which the deep learning model recognizes the signal from the heat maps is obvious. This visualization method obtains a perspective of how the deep learning model handles the signal.

In short, the model has learned the feature space and to focus on special areas after being trained, which is one of the reasons for why the deep learning model performs so well.

\textbf{The fitting of the theoretical and experimental curves with the master equation.} Figure~\ref{theory_and_experiment} shows the fitting of the theoretical and experimental curves with the master equation. If there is no internal or external noise, the master equation fits the spectrum perfectly, as shown in Fig.~\ref{theory_and_experiment}(a). However, when the noise occurs in the experiment, the master equation performs poorly; see Fig.~\ref{theory_and_experiment}(b). Meanwhile, in Fig. 2(e) of the main text, we presented the performance of the deep learning model being trained and tested on the data with biased additional noise. In the last row of Fig. 2(e) of the main text, the model is trained on a training set without additional noise and is tested on a test set with the additional noise that it has never seen before. However, the deep learning model performs well. Once the deep learning model has learned the signal pattern from the training set without additional noise, it extracts the features from the noisy test set even with the noise it has not seen before. Indeed, to further increase the accuracy of the master equation, prior knowledge must be involved to limit the initial values. In contrast, the deep learning model learns to decode the signal by itself without human intervention.

\renewcommand{\thefigure}{S1}
\begin{figure*}[htbp]
\includegraphics[width=1\columnwidth]{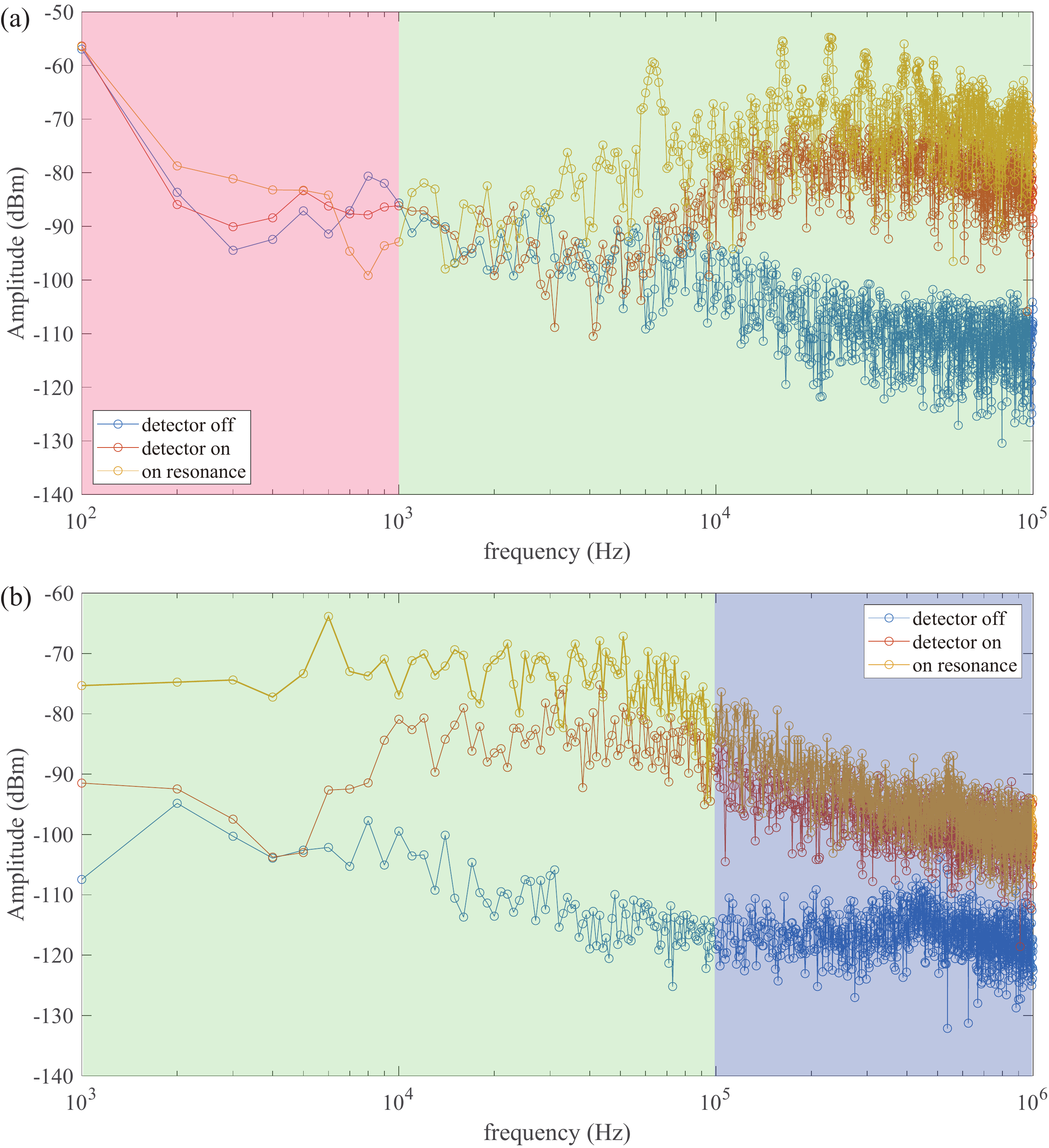} \caption{\textbf{Systematic noise spectra.} The blue curve is the spectrum measured when the differential photodetector was switched off. The red curve is the spectrum measured when the detector was switched on but without a light signal. The yellow curve is the spectrum measured with the EIT configuration but without the microwave signal. The resolution bandwidth and the video bandwidth of the spectrum analyzer are both 10 Hz, and the attenuation is 10. }

\label{spectrum}
\end{figure*}

\renewcommand{\thefigure}{S2}
\begin{figure*}[htbp]
\includegraphics[width=1\columnwidth]{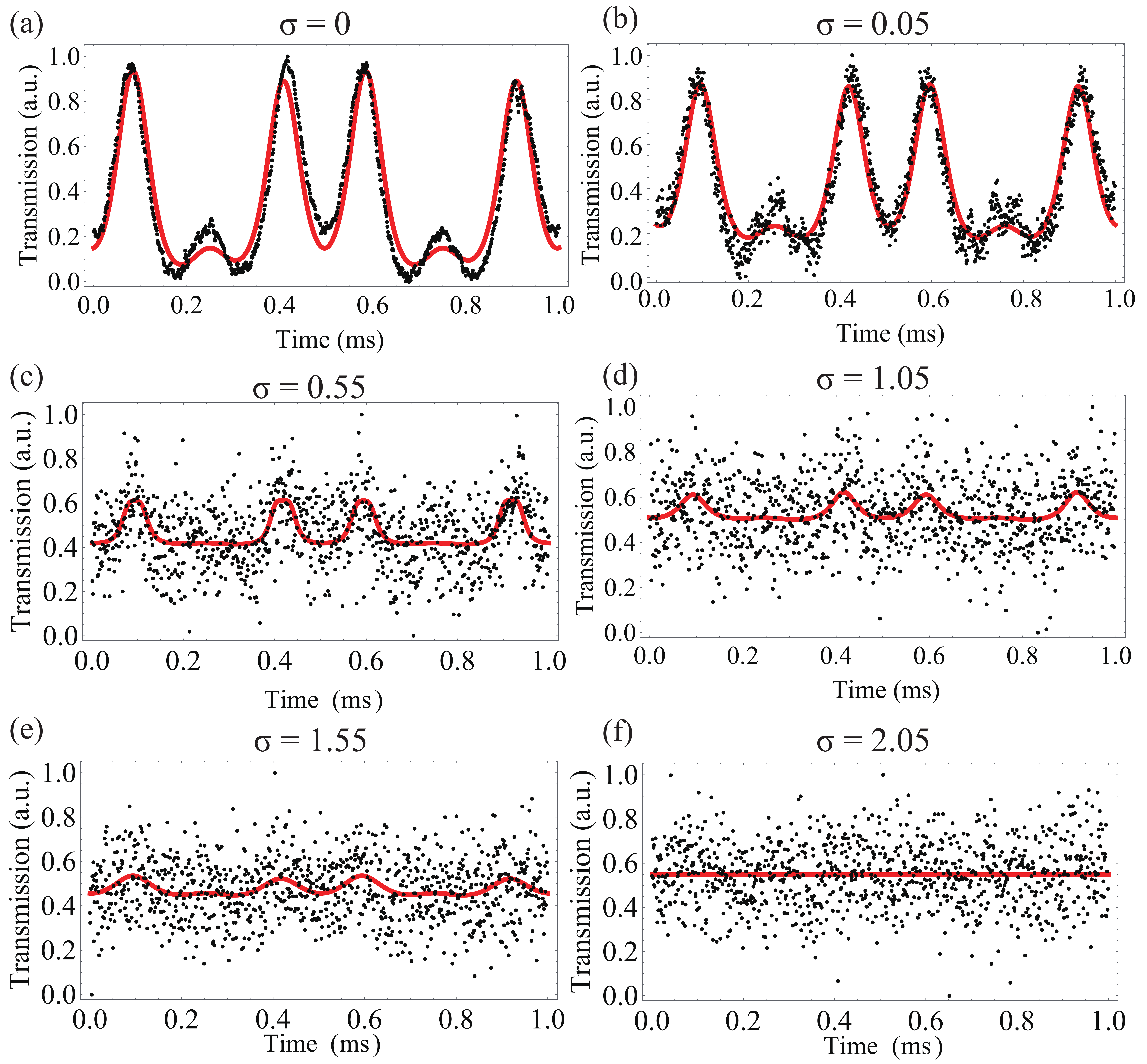} \caption{\textbf{Fitting curve for the master equation on data with additional white noise.} The standard deviation values of the white noise are $\sigma = $ 0 (a), 0.05 (b), 0.55 (c), 1.05 (d), 1.55 (e), and 2.05 (f), respectively. The prediction results are (0, 0, $\pi$, 0), (0, 0, $\pi$, 0), (0, 0, $\pi$, 0), (0, 0, $\pi$, 0), (0, 0, $\pi$, 0), and (0, 0, 0, 0), while the ground truths are both (0, 0, $\pi$, 0).}

\label{spectrum_scale}
\end{figure*}

\renewcommand{\thefigure}{S3}
\begin{figure}[htbp]
\includegraphics[width=1\columnwidth]{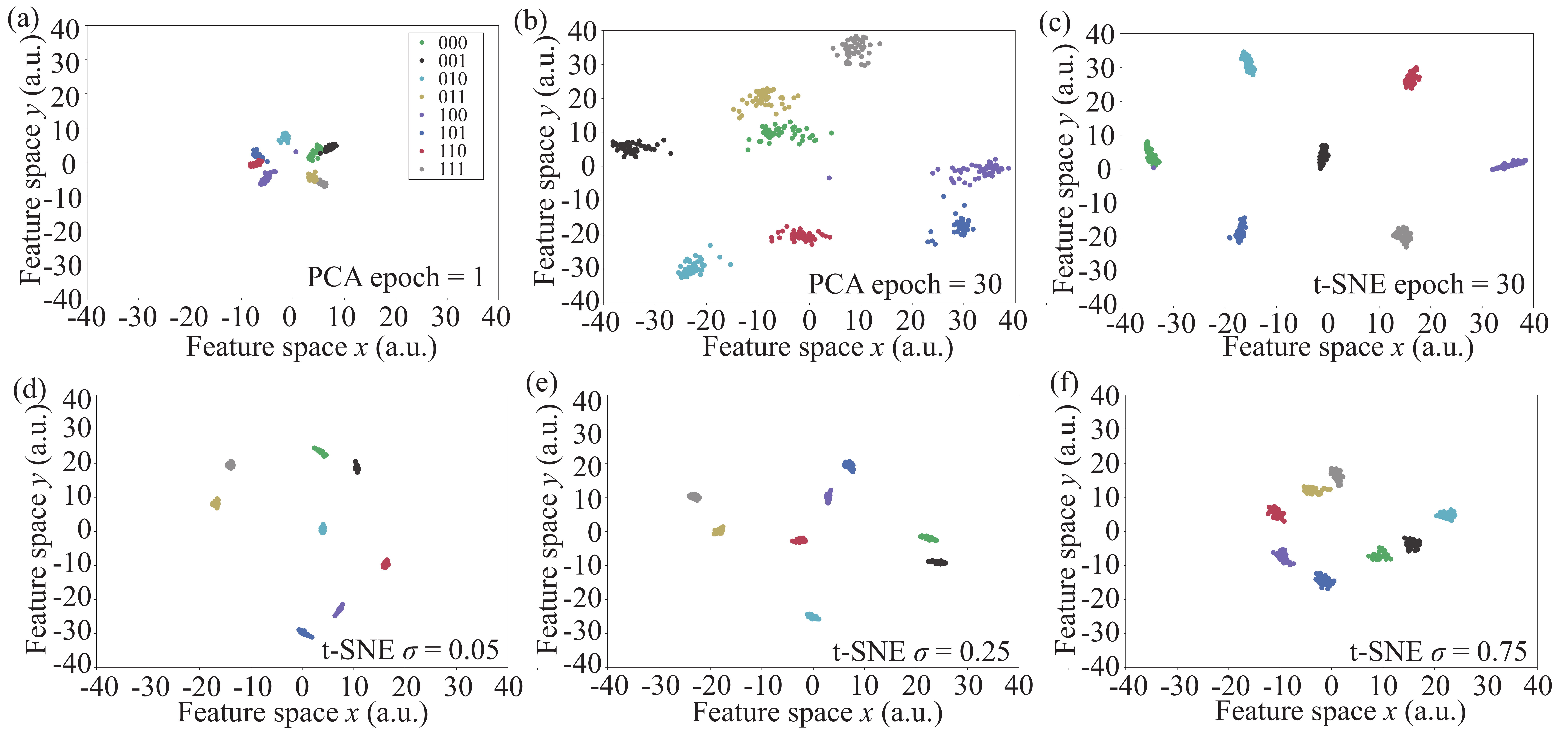} \caption{\textbf{Visualization of the intermediate results of multiple inputs.} Dimensionality reduction methods are adopted. Each colored point represents a signal carrying a 3-bit message, and the same color data points carry the same message. This two-dimensional space is the feature space, where the points carrying the same message are clustered together and those carrying different messages are separated. After 1-epoch (a) and 30-epoch (b) training, the data are fed into the deep learning model and the outputs of the max-pooling layer are handled in principal component analysis. In (c) another method (T-distributed Stochastic Neighbor Embedding) is adopted to deal with the intermediate results of the 30-epoch model. (dšCf) t-SNE results with the standard deviation of additional noise $\sigma = 0.05, 0.75$  and $0.95$. }

\label{pca}
\end{figure}

\renewcommand{\thefigure}{S4}
\begin{figure}[htbp]
\includegraphics[width=1\columnwidth]{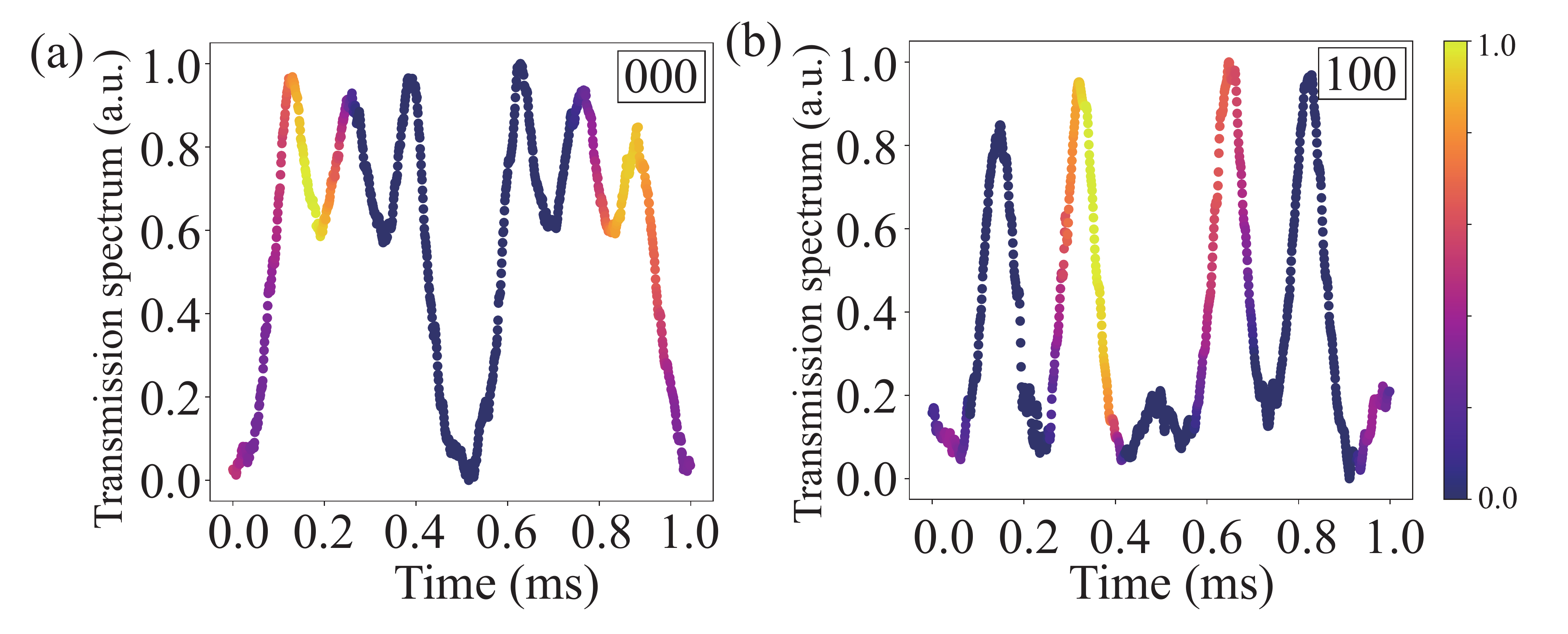} \caption{\textbf{Heat map of the model for singular data.} The color represents how many focuses the model has on the local area of the data. The model recognizes the transmission spectrum via the area with bright color. (a) and (b) are heat maps for the data carrying messages ``000'' and ``100'', respectively.}

\label{heatmap}
\end{figure}

\renewcommand{\thefigure}{S5}
\begin{figure}[htbp]
\includegraphics[width=1\columnwidth]{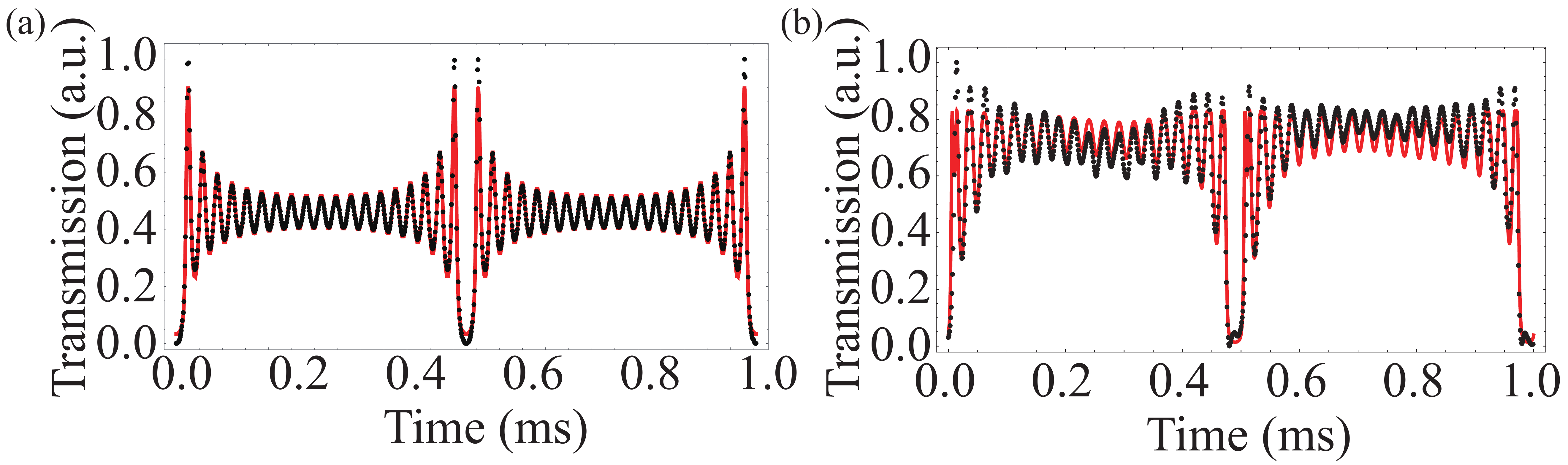} \caption{\textbf{Using the master equation to fit the theoretical curve (a) and experimental curve (b).} The theoretical curve is generated by the master equation. In (a), the master equation fits the data well, whereas in (b), the master equation fits the data poorly.}

\label{theory_and_experiment}
\end{figure}

\clearpage